\documentclass[twocolumn,epjc3]{svjour3}          

\usepackage{amsmath} 

\usepackage{amssymb} 
\usepackage{graphicx}

\usepackage[T1]{fontenc}
\usepackage[latin9]{inputenc}

\allowdisplaybreaks[2]

\newcommand\be{\begin{equation}}
\newcommand\ee{\end{equation}}
\newcommand\ba{\begin{eqnarray}}
\newcommand\ea{\end{eqnarray}}

\newcommand\de{\delta}

\newcommand\al{\alpha}
\newcommand\pa{\partial}
\newcommand\Om{\Omega}

\newcommand\Omk{\Omega_k}
\newcommand\om{\omega}
  
  \newcommand{\call}{{\cal L}}
  \newcommand{\e}{{\rm e}}

\renewcommand{\(}{\left(}
\renewcommand{\)}{\right)}
\renewcommand{\[}{\left[}
\renewcommand{\]}{\right]} 

\newcommand{\lm}{\lambda}
\newcommand{\sig}{\sigma}
\newcommand{\ga}{\gamma}

\newcommand{\bt}{\beta}
\newcommand{\ph}{\phantom{\al}}
\newcommand{\mc}{\mathcal}

\newcommand\Dp{D_{,\phi}}

\newcommand\DY{D_{,Y}}
\newcommand\DYY{D_{,YY}}
\newcommand\Dpy{D_{,\phi Y}}

\newcommand{\pr}{\phi_{,\tilde{r}}}
\newcommand{\prr}{\phi_{,\tilde{r}\tilde{r}}}
\newcommand{\tr}{\tilde{r}}
\newcommand{\tY}{\tilde{Y}}
\newcommand{\trho}{\tilde{\rho}}

\newcommand{\nnb}{\nonumber}

\usepackage{adjustbox}

\journalname{Eur. Phys. J. C}

\begin{document}

\title{Vainshtein mechanism in general purely disformal gravity theory}

\author{Khamphee Karwan\thanksref{e1}$^{,1}$, 
David F. Mota\thanksref{e2}$^{,2}$,
Saksith Jaksri\thanksref{e3}$^{,1,3}$}

\thankstext{e1}{e-mail: khampheek@nu.ac.th}
\thankstext{e2}{e-mail: d.f.mota@astro.uio.no}
\thankstext{e3}{e-mail: saksith@tuta.io}

\institute{ The Institute for Fundamental Study ``The Tah Poe Academia Institute'', Naresuan University,
Phitsanulok 65000, Thailand,\\
$^2$ Institute of Theoretical Astrophysics, University of Oslo, 
P.O. Box 1029 Blindern, N-0315 Oslo, Norway\label{addr2},\\
$^3$ Department of Physics, Faculty of Science, Naresuan University,
Phitsanulok 65000, Thailand\label{addr3}}

\date{Received: date / Accepted: date}

\maketitle

\begin{abstract}
We study a theory of gravity in which the action is a result from the general purely disformal transformation on the Einstein-Hilbert action.
This theory is a sub-class of GLPV theory which is the the generalization of covariant Galileon. 
Nevertheless, we find that the self accelerating solution for the background universe disappears in this theory.
We also find that, for this theory, the Vainshtein mechanism is absent.
 However, the Vainshtein mechanism is not necessary for this theory,
because this theory can nearly mimic the Einstein theory of gravity at all scales inside the Huble radius without this mechanism.
\end{abstract}

\section{Introduction}

One of the most important puzzles in cosmology is the observed accelerated expansion of the late-time universe \cite{SNIa,Planck}.
A possible explanation for this puzzle is that the acceleration of the universe is driven by mysterious form of energy whose pressure is sufficiently negative, called dark energy \cite{Copeland:06}.
On the other hand, the acceleration of the universe can also be a consequence of unknown physics of gravity at cosmic scales.
To achieve the acceleration of the universe, many alternative theories of gravity have been proposed and studied \cite{Clifton:11}.
In the simplest case, the alternative theories of gravity can be constructed by adding scalar degree of freedom to the gravity sector.
These theories belong to the class of scalar tensor theory of gravity \cite{Sotiriou:07}.

An interesting tool for studying the relation among various  theories of gravity is the disformal transformation defined by \cite{Bekenstein:92}
\be
 \bar g_{\mu\nu} = C(\phi,X) g_{\mu\nu} + D(\phi,X)\phi_{,\mu}\phi_{,\nu}\,,
\label{disfull}
\ee
where $X \equiv - \phi_{,\al}\phi^{,\al} / 2$ is the kinetic energy of the scalar field, subscript ${}_{,\mu}$ denotes a partial derivative $\pa_\mu$,
while $C(\phi,X)$ and $D(\phi,X)$ are the coefficients for conformal and disformal transformations respectively.
The above transformation will become the conformal transformation if $D =0$.
Using the conformal transformation with $C = C(\phi)$,
the action for Brans-Dicke theory can be transformed to take the form of the Einstein-Hilbert action in the Einstein frame.
The conformal transformation to the Einstein frame is also possible for  the simple scalar tensor theory of gravity in which the non-minimal coupling is proportional to  $f(\phi) R$ where $f(\phi)$ is an arbitrary function of the scalar field $\phi$ and $R$ is the Ricci scalar.
The physical equivalent between the Einstein frame and the original frame is shown in \cite{Shinji:04}.
However, in order to transform more general scalar tensor theories such as the Horndeski theory \cite{Horndeski,Deffayet:11} and its extensions to the Einstein frame, the disformal transformation is required.
It has been shown in \cite{Bettoni:13} that for suitable coefficients of the Lagrangian the Horndeski action can be transformed to the Einstein frame using the disformal transformation with $C = C(\phi)$ and $D = D(\phi)$.
For this choice of $C$ and $D$,
the form of the Horndeski action is preserved under the disformal transformation.
In the case of the general disformal transformation where $C$ or $D$ depends on the kinetic term $X$,
the disformal transformation can lead to the terms in the action which are beyond the Horndeski theory.
This implies that the theories of gravity obtained from the general disformal transformation may have higher-order time derivative in the equations of motion, and consequently these theories may encounter the Ostrogradski's instability.
However, in some cases, the higher-order time derivative in the equations of motion can reduce to the second order time derivative due to hidden constraints \cite{Zumalacarregui:13}.
an interesting extension for the Horndeski theory, called GLPV theory, has been proposed in \cite{glpv1,glpv2}.
Although the equations of motion for this theory are of third order in derivative in general,
the equations of motion become second order in the flat FLRW universe.
Furthermore, the results from the Hamiltonian analysis indicate that this theory has 1 scalar degree of freedom and 2 tensor degrees of freedom,
and  is free from the Ostrogradski's instability.
It has been shown that each of the non-Horndeski parts of the GLPV action can be separately transformed to subclass of Horndeski action using the disformal transformation with $C = C(\phi)$ and $D = D(\phi,X)$,
but the full GLPV action cannot be obtained by applying the disformal transformation to the Horndeski action \cite{glpv2}.
For this choice of $C$ and $D$,
The structure of the GLPV action is preserved under the disformal transformation \cite{glpv2}.

In addition to instabilities-free, viable theories of gravity are required to recover the Einstein theory in the solar system,
because the predictions from the Einstein theory perfectly satisfy the gravitational experiments in side the solar system.
To recover the Einstein theory, the  fifth force associated to scalar degree of freedom in the gravity sector has to be screened \cite{shaw,clifton,bour}.
For $f(R)$ gravity \cite{Hu:07,Star:07}, the effective mass of the scalar degree of freedom becomes large in high-density regions due to the interaction between scalar degree of freedom and matter,
and therefore the fifth force can be suppressed by the chameleon mechanism \cite{chame,gan}.
The screening of the fifth force can also be at work  due to the non-linear self-interaction of the scalar degree of freedom  through the Vainshtein mechanism \cite{Vainshtein}.
Based on the Vainshtein mechanism, the fifth force can be suppressed for distance smaller than the Vainshtein radius $r_{v}$.
For a static and spherically symmetric background,
the Vainshtein mechanism can work  in Horndeski theory \cite{DeFelice:11,Kase:13}.
However, in a cosmological background, the time variation of the  Newton's constant cannot be suppressed by the Vainshtein mechanism and
the metric potentials are not proportional to inverse distance satisfying the Newtonian  gravity on small scales if $\pa G_5 / \pa X \neq 0$ \cite{Kimura:11}.
The stability of the spherically symmetric screened solutions has been studied in \cite{KNT}.
It has been shown that in the cosmological background, the non-Horndeski pieces in GLPV theory  can lead to a partial breaking of the Vainshtein mechanism inside the compact object \cite{Koba,Sakstein,Mizuno}.
However, for a static and spherically symmetric background, 
the Vainshtein mechanism can work both inside and outside the compact object \cite{anto:15,anto:152}.

In this work, we study cosmology of the disformal gravity theory in which the action is obtained by applying the general purely disformal transformation to the Einstein-Hilbert action.
The evolution of the background FLRW universe and the screening mechanism in this disformal gravity are investigated.
The evolutions for the background universe and the density perturbations for the disformal gravity theory obtained from the disformal transformation with $C = C(\phi)$ and $D=D(\phi)$ have been studied both in the Einstein  frame \cite{David:12,Sakstein:141,vandeBruck:15}
and in the Jordan frame \cite{Sakstein:15}.
It has been shown that the deviation from the Einstein theory due to the disformal coupling
can be suppressed when $\phi$ is slowly varying in time without requirement of the non-linear screening mechanisms \cite{Sakstein:14isa,Sakstein:152}.
Since the disformal coefficient depends solely on the scalar field, the disformal gravity studied in the mentioned works is a subclass of the Horndeski theory.
Here, we consider the other class of disformal gravity theory arisen from the purely disformal transformation in which the disformal coefficient takes more general form, i.e., $D=D(\phi,X)$, but $C = 1$.
Hence, the disformal gravity discussed in this work is the generalization  of the disformal gravity studied in the literature,
such that its action also contains terms which belong to the GLPV theory.
Based on the construction of the disformal gravity,
it is easy to conclude that the disformal gravity obtained from the general disformal transformation, $D = D(\phi,X)$, should be a subclass of the GLPV theory in which the action can be completely transformed to the Einstein-Hilbert form.
The subclasses of the GLPV theory which have such properties cannot be obviously obtained from the full GLPV action by fixing the form of the coefficients in the action.

In sec.~2, the action for this disformal gravity theory is derived by applying the general purely disformal transformation to the Einstein-Hilbert action,
and then we write the resulting action in the form of GLPV action.
in sec.~3, the late time evolution of the FLRW universe for this theory is studied.
The spherically symmetric solutions and the screening mechanism for this theory is considered in sec.~4,
and we  conclude in sec.~5.
The derivation of action for the disformal gravity and its relation with GLPV action are presented in detail in the appendix.

\section{General purely disformal gravity theory}
\label{theory}

In this section, we will derive the action of gravity by applying the general purely disformal transformation on Einstein theory of gravity.
Under the purely disformal transformation the metric tensor is transformed as
\begin{eqnarray}\label{disfrescale}
 \bar g_{\mu\nu} = g_{\mu\nu} + D(\phi,X)\phi_{,\mu}\phi_{,\nu}\,.
\end{eqnarray}
The inverse of the above metric is
\begin{equation}\label{inverse}
\bar{g}^{\mu\nu}= g^{\mu\nu} - \ga^2 D\phi^{;\mu}\phi^{;\nu}\,,
\;\quad \text{ where } \ga^2 \equiv \frac{1}{1 - 2DX}\,,
\end{equation}
One can show that the connections computed from barred metric and original metric are related by \cite{David:12}
\be
\bar\Gamma^{\al}_{\mu\nu} - \Gamma^{\al}_{\mu\nu} \equiv \mathcal{K}^{\al}_{\ph \mu\nu} 
= \bar{g}^{\al\lm}\(\nabla_{(\mu}\bar{g}_{\nu)\lm} - \frac 12 \nabla_{\lm}\bar{g}_{\mu\nu}\)\,.
\ee
Using eqs.~(\ref{disfrescale}) and (\ref{inverse}), we can write the above equation as
\be \begin{split}   
\mathcal{K}^{\al}_{\ph \mu\nu} 
=&
\ga^2 \phi{}^{;\al}\(D{}_{;(\mu} \phi{}_{;\nu)} 
+ D \phi{}_{;\mu}{}_{;\nu} + \frac 12 D \phi{}_{;\mu}\phi{}_{;\nu} \phi{}_{;d1} D{}^{;d1}\) \\
&- \frac 12 D{}^{;\al} \phi{}_{;\mu} \phi{}_{;\nu}\,,
\label{kamn}\end{split}
\ee
where subscript ${}_{;}$ denotes covariant derivative associated to metric $g_{\mu\nu}$.
From the definition of $\mathcal{K}^{\al}_{\ph \mu\nu} $,
one can compute the Ricci scalar using the relation
\ba
\bar{R} &=& \bar{g}^{\bt\nu} \bar{R}^\al_{\phantom{\al}\bt\al\nu}\nnb\\
&=& \bar{g}^{\bt\nu} R^\al_{\ph\bt\al\nu}
 + \bar{g}^{\bt\nu}\nabla_{[\al}\mc K^{\al}_{\ph \nu]\bt} 
+ \bar{g}^{\bt\nu} \mc K ^\al_{\ph \gamma [\al} \mc K^\gamma _{\ph \nu]\bt}\,.
\label{ricci-form}
\ea
From the calculations in the appendix (\ref{riemann-cal}),
we can use the above equation to express the Ricci scalar in a barred frame in terms of the unbarred quantities
Hence, let us consider the action of the form
\ba
S &=& \frac{M_p^2}{2}\int d^4 x \sqrt{-\bar{g}}\, \bar{R}
+ \int d^4x \sqrt{-g}\left(P(\phi,X)\right. \nnb \\ && \left.+ \call_m(g_{\al\bt},\psi)\right)\,,
\label{act-start}
\ea
where $M_p \equiv (8\pi G)^{-1/2}$, $P(\phi,X)$ and $\call_m$ are the Lagrangian density of the scalar field $\phi$ and matter in the unbarred frame.
Using the expression for $\bar{R}$ from eq.~(\ref{ricci-sim1}),
we can write the gravity part of the above action as
\ba
S_g &=& \frac{M_p^2}{2}\int d^4 x \sqrt{-g}\biggl\{
\frac 1{\ga} R - \ga\phi^{;\al}\phi^{;\bt} R_{\al\bt}
\nonumber\\
&&+ \tfrac{1}{2\ga} D{}^{;\al} \phi{}_{;\al}{}_{;\om} \phi{}^{;\om}
+\gamma \Bigl[2 X D{}^{;\al}{}_{;\al} + D \phi{}^{;\om}{}_{;\om}{}_{;\al} \phi{}^{;\al}\nonumber\\
&& + D \phi{}^{;\al}{}_{;\al} \phi{}^{;\om}{}_{;\om} + \phi{}_{;\al} D{}^{;\al} \phi{}^{;\om}{}_{;\om} -  D \phi{}^{;\al} \phi{}_{;\al}{}^{;\om}{}_{;\om} 
+
\nonumber\\
&&
+ \phi{}^{;\al} D{}_{;\al}{}_{;\om} \phi{}^{;\om} -  D{}^{;\al} \phi{}_{;\al}{}_{;\om} \phi{}^{;\om} -  D \phi{}_{;\al}{}_{;\om} \phi{}^{;\al}{}^{;\om}\Bigr]
\nonumber\\
&&
+ \tfrac{1}{2} \gamma^3 \Bigl[
4 X^2 D{}_{;\al} D{}^{;\al} 
+ 2 \phi{}_{;\al} D{}^{;\al} \bigl(\phi{}^{;\om}{}_{;\om} + X \phi{}_{;\om} D{}^{;\om} 
\nonumber\\
&& -  D \phi{}^{;\om} \phi{}_{;\om}{}_{;\lm} \phi{}^{;\lm}\bigr) 
-  \phi{}^{;\om} \bigl((3 + 4 D^2 X^2) D{}^{;\al} \phi{}_{;\al}{}_{;\om} 
\nonumber\\
&&
+ 4 D^2 \phi{}^{;\al} (\phi{}_{;\al}{}_{;\om} \phi{}^{;\lm}{}_{;\lm} -  \phi{}_{;\om}{}_{;\lm} \phi{}_{;\al}{}^{;\lm})\bigr)\Bigr]
\biggr\}\,.
\label{act0}
\ea
The action for the gravity resulting from applying the general transformation, in which $C = C(\phi,X)$ and $D = D(\phi,X)$, to the Einstein-Hilbert action has been derived in \cite{Zumalacarregui:13}.
However, in order to obtain the above action from the action in \cite{Zumalacarregui:13}, non-trivial integrations by parts  are needed.
The relations between the coefficients of the Horndeski theory and the GLPV theory through the general purely disformal transformation given in eq.~(\ref{disfrescale}) have been discussed in \cite{glpv2} for the unitary gauge.
Starting from the above action, we perform several integration by parts shown in appendix (\ref{byparts}) to obtain the simplified action as
\ba
S_g &=& \frac{M_p^2}{2}\int d^4 x \sqrt{-g}\biggl\{\,\,
\frac 1{\ga} R 
- \gamma D \Bigl[
\(\Box \phi\)^2
- \phi{}_{;\al}{}_{;\bt} \phi{}^{;\al}{}^{;\bt}\Bigr]
\nonumber\\
&&+\ga\Bigl[
D{}^{;\al} \phi{}_{;\al}{}_{;\om} \phi{}^{;\om}
-
\phi{}_{;\al} D{}^{;\al} \phi{}^{;\om}{}_{;\om} 
\Bigr]
\biggr\}\,.
\label{act35}
\ea
It is clear that the above  action is not the Horndeski action.
Hence, to ensure that this theory of gravity is free of ghost, we will transform this action in to the GLPV form.
Let $G_4(\phi, X) \equiv M_p^2 / (2 \gamma)$,
and defined
\be
Y \equiv - 2 X = \phi_{,\al}\phi^{,\al}\,,
\label{def-y}
\ee
we can insert the action (\ref{act35}) in to the action (\ref{act-start}) ,
and write the resulting action in the GLPV form \cite{glpv1,glpv2,anto:15} as 
\be
S = \int d^4 x \sqrt{-g} \sum^{4}_{i=2}L_i
+\int d^4 x \sqrt{-g}\,\call_m(g_{\al\bt},\psi)\,,
\label{act-glpv}
\ee
where 
\ba
L_{2} &=& A_2(\phi,Y) + Y C_{3,\phi}
\,,
\label{L2} \\
L_{3} &=& \left( C_{3}+2YC_{3,Y} \)\Box\phi\,,  
\label{L3} \\
L_{4} &=& B_{4} R - \frac{B_{4}+A_{4}}{Y}\[(\Box\phi)^{2}-\nabla ^{\mu}\nabla^{\nu}\phi \nabla_{\mu}\nabla_{\nu}\phi\]
  \nonumber \\
&&+ \frac{2\left( B_{4}+A_{4}-2Y B_{4,Y}\)}{Y^{2}} \(\nabla^{\mu}\phi \nabla^{\nu}\phi \nabla_{\mu}\nabla_{\nu}\phi\,\Box \phi
\right.
\nonumber \\
& &\left.
-\nabla^{\mu}\phi \nabla_{\mu}\nabla_{\nu}\phi \nabla_{\sigma}\phi \nabla^{\nu}\nabla^{\sigma}\phi \)\,.
\label{L4} 
\ea
Here, $A_2 \equiv P(\phi, Y)$ is the Lagrangian of scalar field in eq.~(\ref{act-start}).
It follows from the appendix (\ref{glpv}) that for our case, we have
\ba
C_3 &=& - \frac{M_p^2}{2} \frac 12 \int \ga \Dp dY\,,
\qquad
B_4 = \frac{M_p^2}{2} \frac 1{\ga}\,,
\nonumber\\
A_4 &=& -G_4 + 2 Y G_{4,Y}- Y^2 F_4
= - \frac{M_p^2}2 \ga\,.
\label{c3a4}
\ea
Using the relation between $C_3$ and $A_3$ in \cite{glpv1,anto:15,anto:152}, one can show that
\be
A_3 = 2 (-Y)^{3/2}   C_{3,Y}
- 2 \sqrt{-Y}B_{4,\phi}
= 0\,. ,\\
\label{a3}
\ee
In the following sections, we will set $M_p^2 = 1$ for convenience.

\section{Background evolution}
\label{acc}

The evolution equations for the background universe can be obtained by supposing that the field $\phi$ is homogeneous, i.e., $\phi = \phi(t)$,
and using the FLRW metric given by
\be
ds^2 = - n(t)^2 dt^2 + a(t)^2 \delta_{ij}dx^i dx^j\,,
\label{ds-flrw}
\ee
where $\de_{ij}$ is the Kronecker delta.
Varying the action (\ref{act-glpv}) with respect to $n(t)$ and $a(t)$,
and then setting $n(t) = 1$, we respectively get 
\ba
0 &=& 
\left( A_2 -2 Y A_{\text{2,Y}}\) - \rho_m 
+ 3 H^2 \ga \frac{1 - Y^2 \DY}{1 + D Y}\,,
\\
0 &=&
- H \ga^3 \dot\phi \( Y \Dp - 2 \(D + Y \DY\) \ddot\phi\)
+ \ga\(2 \frac{\ddot{a}}{a}  + H^2\)
\nonumber\\
&&
+ A_2 + p_m \,,
\label{addot}
\ea
where a dot denotes a derivative with respect to time, $H = \dot a / a$ is the Hubble parameter, $\rho_m$ and $p_m$ are the energy density and pressure of matter respectively.
Since the disformal gravity considered in this work is a sub class of the GLPV theory which is the covariantized Galileon theory \cite{anto:1503},
we first check whether the acceleration of the universe can be driven by the kinetic terms of scalar field as in the Galileon theory\cite{Sami,DT10}.
In the flat FLRW background, we have
$\ga = 1 / \sqrt{1 - D\dot\phi^2}$,
so that $D \dot\phi^2$ should lie within the range $(-\infty, 1)$.
In addition, it follows from the above equations that $\ga$ should be unity during matter dominated epoch and should be larger than unity during the acceleration of the universe.
Hence, $0\leq D \dot\phi^2 < 1$ throughout the evolution of the universe.
To study how the universe can be accelerated,
we use $Y = - {\dot\phi}^2$ to write eq.~(\ref{addot}) as
\be
\frac{\ddot a}{a} = - \frac 1{2\ga}\(\ga H^2 + A_2 + p_m +2 H \dot{\ga}\)\,.
\label{addot1}
\ee
Since $\ga$ increases in time,
the contribution from the $\dot\ga$-term cannot lead to an accelerated expansion of the universe.
Hence, the accelerated expansion of the universe can be achieved only if the pressure of the scalar field is sufficiently negative, i.e. $A_2 \equiv p_\phi < -\rho_\phi/3$ for $\ga \sim 1$.
Here, $\rho_\phi \equiv 2 Y A_{\text{2,Y}} - A_2$.
Therefore, for the disformal gravity considered here, the accelerated expansion of the universe cannot be driven by kinetic terms of the scalar field, i.e., self accelerating solution does not exist.

\begin{figure}[h]
 \centering
 \includegraphics[scale=0.68,keepaspectratio=true]{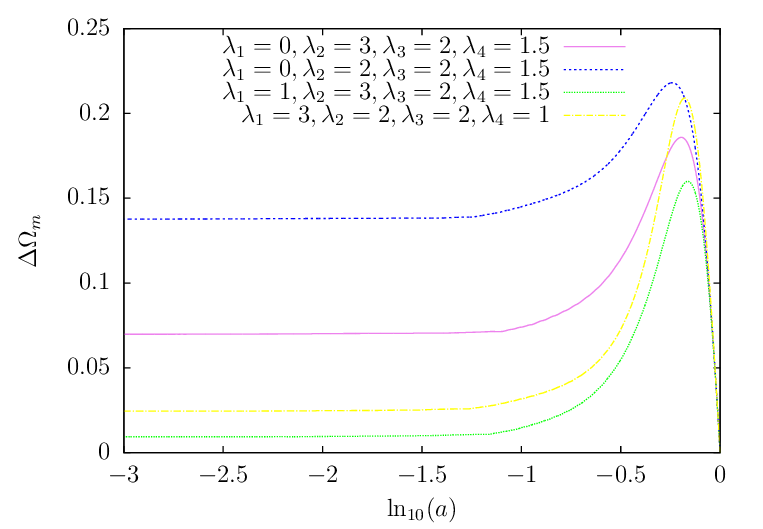}
 \caption{\label{fig1}
 the different density parameter $\Delta\Om_m$ as a function of $\log_{10}a$ for various values of $\lambda_1$, $\lambda_2$, $\lambda_3$ and $\lambda_4$.
}
\end{figure}

\begin{figure}
\centering
\includegraphics[height=0.34\textwidth
,angle=0,keepaspectratio=true]{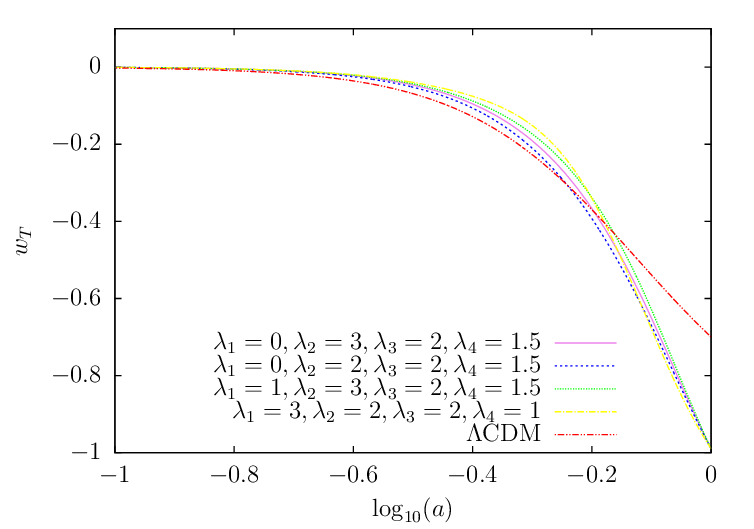}
\caption{\label{fig2}
 The equation of state parameter $w_T$ as a function of $\log_{10}a$ for various values of $\lambda_1$, $\lambda_2$, $\lambda_3$ and $\lambda_4$.
}
\end{figure}

To illustrate how  the background universe evolves at late time for disformal gravity theory,
we solve evolution equations for the background  universe numerically.
Variation of the action with respect to $\phi$ yields 
\ba
0 &=&
\ddot\phi \bigg[A_{\text{2,Y}}+2 Y A_{\text{2,YY}}
+ \frac 32 H^2 \ga^5 \big[
D  \left( 1\right. - Y^2 \DY
\nonumber\\
&&
+ 2 Y^3 \left. \DYY\right)
- 2 Y D^2 
\big]\bigg]
+ Y (5 \DY - 3 Y^2 \DY^2  
\nonumber\\
&& + 2 Y \DYY)
\nonumber\\
&&
+3 H \dot\phi \left(A_{\text{2,Y}}
- \ga^3 Y \(D + Y \DY\)\(\frac 12 H^2 + \frac{\ddot{a}}a\)\)
\nonumber\\
&&
+ \frac{1}{2} \bigg(A_{\text{2,} \phi }-2 Y A_{\text{2,Y} \phi } 
+ \frac 32 H^2\ga^3\bigg[
3 Y^2 \Dp \frac{D + Y \DY}{1 + D Y} 
\nonumber\\
&&- 2 Y^2 \Dpy
-Y \Dp
\bigg]
\bigg)\,.
 \ea
For concreteness, we choose the disformal coupling of the form
\be
D \equiv M^{-4 \lm_2 - 4} \e^{-\lm_1 \phi} (- Y)^{\lm_2}\,,
\label{disf-mod}
\ee
and choose $A_2$ as
\be
A_2 \equiv \frac 12 M_k^{4 -4 \lm_3} (- Y)^{\lm_3} - M_v^4 \e^{- \lm_4 \phi}\,,
\label{a2-mod} ) 
\ee
Here, $M$, $M_k$ and $M_v$ are the constant parameter with dimension of mass,
while $\lm_1$, $\lm_2$, $\lm_3$ and $\lm_4$ are the dimensionless constant parameters.
For the homogeneous and isotropic universe, $Y = - \dot\phi^2$,
and therefore the field $\phi$ may be classified as a phantom field when the kinetic term in $A_2$ is proportional to $Y^{\lm_3}$.
We choose the above form of  $A_2$ because this form can be easily reduced to the canonical form,
and as discuss above, the potential term of the scalar field is needed to drive an accelerated expansion of the universe.
The above form of the disformal coefficient $D$ is chosen
because this form is the simplest form that can be used to study the influence of the kinetic-dependent disformal coefficient.
For this choice of $D$ and $A_2$, the equations of motion become
\ba
0 &=&
3 H^2\(\gamma^3 + \gamma \left(\gamma ^2-1\right) \lambda_2\)
\nonumber\\
&& -\frac{1}{2} \(2\lm_3 - 1\) M_k^{4 - 4\lm_3} (\dot\phi)^{2 \lm_3}  - M_v^4 \e^{-\lm_4 \phi}
-\rho_m
\,,
\label{dis:00}\\
0 &=&
2 \gamma  \frac{\ddot{a}}{a}
- \gamma^3 H \dot\phi \left(\lambda_1-2 D \left(\lambda_2+1\right) \ddot\phi\right)
+ \gamma  H \left(H+\lambda_1 \dot\phi\right)\nonumber\\
&&
+\frac 12 M_k^{4 - 4\lm_3} (\dot\phi)^{2\lm_3} 
- M_v^4 \e^{-\lm_4 \phi}
+ p_m
\,,
\label{dis:ii}\\
0 &=&
+\ddot\phi \Big(M_k^{4 - 4\lm_3} \lm_3 (2 \lm_3-1) (\dot\phi)^{2\lm_3}-3 \gamma^3 D H^2 \left(\lambda_2+1\right) Y \nonumber\\
&& \times
 \left(\lambda_2 \left(3 \gamma^2 D Y-2\right)+3 \gamma^2 D Y-1\right)\Big)
\nonumber\\
&&
+3 H M_k^{4 - 4\lm_3} \lm_3 (\dot\phi)^{2\lm_3} \dot\phi
+Y M_v^4 \lm_4 \e^{-\lm_4 \phi}\nonumber\\
&&+\frac Y{2} \bigg(
- 3 \gamma^3 D H  \Big(H \lambda_1 Y\big( \lambda_2 \left(3 \gamma^2 D Y-2\right)
 \nonumber\\
&&
+3 \gamma ^2 D Y - 1\big)
-2 \left(\lambda _2+1\right) \left(3 H^2+2 \dot{H}\right) \dot\phi\Big)
\bigg)
\,.
\label{dis:kg}
\ea
Substituting $M_v^4 \e^{-\lm_4 \phi}$ from eq.~(\ref{dis:00}) into eq.~(\ref{dis:kg}),
we can write eq.~(\ref{dis:kg}) as
\ba
\phi'' &=&
 \frac{H'}H \phi'
\Big[2 \big(9 \gamma  \left(\gamma ^2-1\right)^2 \left(\lambda _2+1\right){}^2+3 \gamma  \left(\gamma ^2-1\right)
\nonumber\\
&& \times \left(2 \lambda _2+1\right) \left(\lambda _2+1\right)-2 \lambda _3^2 \Omega _k+\lambda _3 \Omega _k\big)\Big]^{-1}
\nonumber\\
&& 
\begin{split}&
\times \phi ' \bigg( 3 \gamma ^3 \Big(2 \lambda _2 \big(2 \frac{H'}{H}+2 \lambda _1 \phi '+\lambda _4 \phi '+3\big)\\
&
+4 \frac{H'}{H}+5 \lambda _1 \phi '+2 \lambda _4 \phi '+6\Big) \\
&-3 \gamma  \Big(\lambda _2 \big(4 \frac{H'}{H}+\lambda _1 \phi '+2 \lambda _4 \phi '+6\big)+4 \frac{H'}{H}\\&+2 \lambda _1 \phi '+6\Big)+\lambda _4 \left(\Omega _k-6 \Omega _m\right) \phi ' 
\\
&
-2 \lambda _3 \Omega _k \left(\lambda _4 \phi '+3\right) -9 \gamma ^5 \lambda _1 \left(\lambda _2+1\right) \phi ' \bigg)
\,,
\end{split}
\label{flrw:kg}
\ea
where a prime denotes a derivative with respect to $N = \ln a$,
$\Omk \equiv M_k^{4 - 4\lm_3} (H\phi')^{2\lm_3}/ H^2$,
$\Om_m = \rho_m / 3H^2 = \Om_m^0 \e^{-3 N} / (H^2 / H_0^2)$, $H_0$ and $\Om_m^0$ are the present value of the Hubble parameter and $\Om_m$ respectively.
The function $H' / H$ can be computed by combining eq.~(\ref{dis:ii}) with eq.~(\ref{dis:00}) and setting $p_m =0$, so that we get
\ba
\frac{H'}H  &=&
\Big[2 \gamma  \big(3 \gamma ^5 \left(\lambda _2+1\right){}^2-3 \gamma ^3 \left(\lambda _2+1\right)\nonumber\\
&& -3 \gamma  \lambda _2 \left(\lambda _2+1\right)+\left(1-2 \lambda _3\right) \lambda _3 \Omega _k\big)\Big]^{-1}\times \nonumber\\
&& \bigg[ 27 \gamma ^8 \left(\lambda _2+1\right){}^3-9 \gamma ^6 \left(\lambda _2+1\right){}^2 \left(7 \lambda _2+6\right)
\nonumber\\
&& 
 +9 \gamma ^4 \left(\lambda _2+1\right){}^2 \left(5 \lambda _2+3\right)-9 \gamma ^2 \lambda _2 \left(\lambda _2+1\right){}^2 \nonumber\\
&&
-9 \gamma ^5 \left(\lambda _2+1\right){}^2 \left(\lambda _3 \Omega _k+3 \Omega _m\right)+3 \gamma ^3 \left(\lambda _2+1\right) \nonumber\\
&&\times \big(2 \left(2 \lambda _2-\lambda _3+2\right) \lambda _3 \Omega _k+3 \left(4 \lambda _2+5\right) \Omega _m\big)\nonumber\\
&&
-3 \gamma  \left(\lambda _2+1\right) \big(\left(\lambda _2-2 \lambda _3+1\right) \lambda _3 \Omega _k\nonumber\\
&&+3 \left(\lambda _2+2\right) \Omega _m\big)+\lambda _3 \left(2 \lambda _3-1\right) \Omega _k \left(\lambda _3 \Omega _k 
+3 \Omega _m\right)\nonumber\\
&&
+\left(\gamma ^2-1\right) \gamma  \phi ' \times\Big( \lambda _1 \left(1-2 \lambda _3\right) \lambda _3 \Omega _k\nonumber \\
&&+\left(\lambda _2+1\right) \lambda _4 \big[6 \gamma ^3+6 \left(\gamma ^2-1\right) \gamma  \lambda _2
\nonumber \\
&&-2 \lambda _3 \Omega _k
+\Omega _k-6 \Omega _m\big]\Big)\bigg]
\,.
\label{hp2h2}
\ea
Setting $\Om_m^0 = 0.3$, $w_T = -0.97 (1 - \Om_m^0) = -0.68$ at present and $M^2 = M_k^2 = M_v^2 = M_p H_0$, where we have restored $M_p$ in this relation to avoid confusion and $w_T \equiv - 2 \dot H / (3 H^2) - 1$,
we numerically solve eqs.~(\ref{flrw:kg}) and (\ref{hp2h2}) by making an integration from the present to the past of the universe,
and plot the evolution of $\Delta\Om_m \equiv (\Om_m - \Om_m^\Lambda)/ \Om_m^\Lambda$ and $w_T$ in figs.~(\ref{fig1}) and (\ref{fig2}).
Here, $\Om_m^\Lambda$ is the density parameter of matter computed from $\Lambda$CDM model by setting $\Om_m^\Lambda = 0.3$ at present.

%
It follows from the plots that the evolutions of $\Om_m$ and $w_t$ for the disformal model mimics that evolutions for $\Lambda$CDM model.
The evolution of the universe will closely mimic the $\Lambda$CDM model if $\lm_3 =1$, so that it is not plotted.
In the numerical integration, $\ga \to 1$ because $-DY \to 0$ during matter dominated epoch,
while $\ga$ becomes larger than unity at late time when $-DY = D \dot\phi^2$ significantly increases from small value.
According to the numerical results, $- DY$ is always smaller than unity throughout the evolution of the universe,
and $\dot\phi / H < 1$, i.e., the field slowly evolves in time compared with the expansion rate of the universe.

\section{Spherically symmetric solutions and screening mechanism}
\label{sec:vain}

\subsection{Spherically symmetric static background}
\label{sub1}

To investigate the screening  mechanism in the disformal gravity,
we first study the solutions in the spherically symmetric static background.
Since the theory of gravity considered in this work is the sub class of GLPV theory,
We study the screening mechanism in this theory based on the analysis in \cite{anto:15}.
In order to study behavior of gravity in the spherical static background, we write the line element in the form
\be
ds^{2}=-\e^{2\Psi(r)}dt^{2}+\e^{2\Phi(r)}dr^{2}
+r^{2} \(d\theta^{2}+\sin^{2}\theta\,d\varphi^{2}\)\,.
\label{line}
\ee
Vary the action (\ref{act-glpv}) with respect to the metric tensor, we get \cite{anto:15}
\ba
&&
\frac{2\e^{-2\Phi}\ga}{r}\frac{1 - Y^2 \DY}{1 + Y D}\Phi'
\nonumber \\
&&+ A_2 
+ {\cal C}- \frac{\ga}{r^2} \(\e^{-2\Phi}- \frac{1}{\ga^2}\)
=\rho_m\,,\label{eq:00}\\
&&
\frac{2\e^{-2\Phi}\ga}{r}\frac{1 - Y^2 \DY}{1 + Y D}\Psi'
- A_2 +2 Y A_{2,Y} 
 \nonumber \\
&&+\frac{\ga}{r^2} \bigg(\e^{-2\Phi}- \frac 1{\ga^2}\bigg)\frac{1 - Y^2 \DY}{1 + Y D}
=p_m\,,
\label{eq:11}\\
&&
\e^{-2\Phi}\ga \( \Psi''+{\Psi'}^2\) 
+ \( \frac{\e^{-\Phi}\ga}{r}-\frac12 {\cal C} r \)\Psi'  
\nonumber \\
&&
- \frac{\e^{-2\Phi}\ga}{r}\frac{1 - Y^2\DY}{1 + YD}\(1+r\Psi'\)\Phi' 
- A_2 - \frac12 {\cal C}
=p_m\,,\label{eq:22}
\ea
where, in this  section, a prime denotes derivative with respect to $r$, $Y = \e^{-2\Phi} \phi'^2$ for this consideration due to the static and spherical assumptions, and
\be
{\cal C} \equiv - \frac{2\e^{-2\Phi} \phi' \left(\ga_{,\phi} + 2\e^{-2\Phi}\phi'' \ga_{,Y} \right)}{r}\,.
\label{Cidef}
\ee
The conservation equation for the matter yields
\be
p_m'+\Psi' \left( \rho_m+p_m \right)=0\,.
\label{eq:33}
\ee
Inserting $\Phi'$ and $\Psi'$ from eqs.~(\ref{eq:00}) and (\ref{eq:11}) into eq.~(\ref{eq:22}) and considering the weak field limit,
we  obtain
\be
\ga \Box\Psi
+ \frac 12 \frac{1 + YD}{1 - Y^2\DY}\(2 Y A_{2,Y} - A_2\)
- \frac 12 A_2 
= 1/2 \rho_m\,,
\label{boxpsi}
\ee
where $\Box\Psi \equiv d^2\Psi / dr^2 + (2/r) d\Psi / dr$, and we have set $p_m =0$.
In the weak field limit, i.e., $|\Phi| \ll 1, |\Psi| \ll 1$,
we suppose that the main contributions to the evolution equations
are of the order of $\Phi$ and $\Psi$
to ensure that the equations of motion will satisfy Einstein theory in the solar system \cite{Kase:13,anto:15}.

Comparing with the analysis in \cite{DeFelice:11},
one can see that eq.~(\ref{boxpsi}) has no contribution from $\Box\phi$.
This suggests that the effective gravitational constant in disformal gravity is the same as that in Einstein theory.

In order to check how much the post-Newtonian parameter
$\Gamma \equiv |1 + \Phi / \Psi|$ deviates from zero,
we write eqs.~(\ref{eq:00}) and (\ref{eq:11})
in the weak field limit as
\ba
&&
\frac{2\ga}{r}\Phi'
+ A_2 + \frac{\ga^3}{r} \phi'\(Y \Dp + 2 \phi'' (D + Y \DY)\) 
\nonumber \\
&&
- \frac{\ga}{r^2} \(-2\Phi - Y D\)
- \rho_m = 0\,,
\label{eq:00w}\\
&&
\frac{2\ga}{r}\Psi'
-A_2 +2 Y A_{2,Y} 
-\frac{\ga}{r^2} \(2\Phi + Y D\)
= 0\,
\label{eq:11w}
\ea
Influences of the scalar field on $\Phi$ and $\Psi$ in the above equations
can  be estimated  by studying how $\phi'$ and also $\phi$ depend on $r$.
The equation of motion for the field $\phi$ can be computed by differentiating eq.~(\ref{eq:11}) with respect to $r$,
and then eliminate $\Psi''$, $\Psi'$ and $\Phi'$ terms in the result using eqs.~(\ref{eq:00}) -- (\ref{eq:33}).
In the weak field limit, we can write the equation of motion for $\phi$ as \cite{anto:15,anto:152}
\be
\phi'' + \frac 2r \phi' = \frac{\phi' r}{2 \ga} \rho_m + \mu\,,
\label{boxp}
\ee
where
\ba
\mu &\equiv&
\bigg[
\frac{r^2}{2} \left(A_{2,\phi }-2 Y A_{2,\phi Y}\right)
+ r 4 Y^{3/2} A_{2,YY}- \frac{\gamma^5}{4} Y^2 \times \nonumber \\
&&
big(\left(Y (2-D Y) D_{,Y}+3 D\right) D_{,\phi }+2 D Y (D Y+1) D_{,\phi Y} \big)
%
\nonumber\\
&&
+ \frac{\gamma^5}r  Y^{3/2} \Big( D (2 Y^2 (D Y+1) D_{,YY}+3 D)
\nonumber \\ &&+Y^2 (2-D Y) D_{,Y}^2
+3 D Y (D Y+3) D_{,Y} \Big)
\bigg]\[r \beta\]^{-1},
\label{mu}\\
\beta &\equiv&
\(2 Y A_{2,YY}+A_{2,Y}\)r
+\frac{\gamma^5}{2r} \Big(Y \big(D (2 Y^2 (D Y+1) D_{,YY}\nonumber \\
&&+3 D)+Y^2 (2-D Y) D_{,Y}^2
+3 D Y (D Y+3) D_{,Y}\big)\Big)\,.
\label{beta}
\ea
The solutions for eq.~(\ref{boxp}) are conveniently obtained by  replacing the distance $r$ by $\tilde{r} \equiv r H_0$,
so that we can write eq.~(\ref{boxp}) as
\be
\prr + \frac 2{\tr} \pr - \frac{\pr}{2\ga\tr} \tr^2 \trho_m = \tilde\mu\,,
\label{boxph0}
\ee
where ${}_{,\tr}$ denotes derivative with respect to $\tr$,
$\trho_m \equiv \rho_m / H_0^2$ which equals to $3\Om_m^0 \lesssim 1$ for the 
background density while becomes much larger than unity inside the gravitational source.
However, since the radius of the sun and the Milky Way are respectively $\tr_s \sim 10^{-18}$ and $\tr_g \sim 10^{-5}$,
we have $\tr^2 \trho_m \lesssim 10^{-6}$ inside the sun and Milky Way if $\trho_m$ is computed from the mean energy density of these objects.
Outside the Milky Way, one may assume that $\trho_m \sim 3\Om_m^0$, so that $\tr^2 \trho_m \sim \tr^2$.
Hence, if we suppose that $r = 0$ at center of the sun,
the third term on the LHS of eq.~(\ref{boxph0}) is negligible compared with the second term as long as $\tr \lesssim 1$.
This implies that the gradient of $\phi$ is not significantly sourced by $\rho_m$.

To study how the non-linear terms of $\pr$ in $\tilde\mu$ influence behavior of solution for eq.~(\ref{boxph0}),
we use the condition $\pr \to 0$ when $\tr \to 0$.
This condition suggests that, when $\tr \to 0$, $\pr \propto \tr^p$ where $p > 0$.
Based on this property of $\pr$, the dominant terms in the expression for $\tilde\mu$ near $\tr =0$ can be written for the case where the first term on the RHS of eq.~(\ref{beta}) decreases to zero slower than the other terms when $r \to 0$ as
\ba
\tilde\mu &\simeq &
\frac 1{2 H_0^2} \frac{A_{2,\phi }-2 Y A_{2,\phi Y}}{
A_{2,Y} + 2 Y A_{2,Y}}
+ \frac{4 \pr}{\tr}\frac{Y A_{2,YY}}{
A_{2,Y} + 2 Y A_{2,YY}}\,,
\nonumber\\
{}&&
\label{mu1}
\ea
where we have supposed that $D$ is a polynomial function of $Y$.
For the opposite case, $\tilde\mu$ becomes
\ba
\tilde\mu &\simeq &
\frac{\tr^2\tY}{H_0^2} \frac{A_{2,\phi }-2 \tY A_{2,\phi \tY}}{
\(4\lm_2^2+7\lm_2 +3\)Y^2D^2}
%
%
- \frac{D_{,\phi}}{2D} \frac{\tY}{\lm_2 +1}
%
+ \frac{2}{\tr} \pr\,,
\nonumber\\
{}&&
\label{mu2}
\ea
where eq.~(\ref{disf-mod}) has been used and $\tY \equiv Y / H_0^2$.

Let $A_2$ be decomposed as $A_2 = A_K(Y) + A_V(\phi)$, which its concrete form is given by eq.~(\ref{a2-mod}).
Inserting $\tilde\mu$ from eqs.~(\ref{mu1}) and (\ref{mu2}) into eq.~(\ref{boxph0}),
we respectively get
\ba
{}&&
u_1' +\frac{2 u_1}{\tr}
= (-1)^{\lm_3} \frac{\lm_4}{\lm_3}\,,
\label{u1p}\\
{}&&
u_2' 
- \frac{\lm_1}{2} \frac{4 \lm_2 +3}{\lm_2 +1} u_2^{\displaystyle{\frac{4\lm_2 + 4}{4\lm_2 +3}}} 
=
- \frac{\lm_4\tr^2}{\lm_2 + 1}\,,
\label{u2p}
\ea
where $u_1 \equiv \pr^{2\lm_3 - 1}$,
$u_2 \equiv \pr^{4\lm_2 +3}$ and
$\e^{-\lm_4\phi} \sim 1$ is assumed.
Imposing the condition $u_1, u_2 \to 0$ when $\tr \to 0$,
The solutions for the above equations are
\ba
u_1 &=& (-1)^{\lm_3} \frac{\lm_4 \tr}{2 \lm_3}\,,
\,\,\mbox{so that}\,\,\pr \sim (-1)^{\lm_3} \tr^{1 / (2\lm_3 - 1)}\,,
\label{u1s}\\
u_2 &=& - \frac{\lm_4\tr^3}{3\lm_2 + 3}\,,
\,\,\mbox{so that}\,\, \pr \sim - \tr^{1/(4\lm_3 + 3)}\,.
\label{u2s}
\ea
In the case where the first term on the RHS of eq.~(\ref{beta}) decreases slower than the other terms when $\tr \to 0$,
we have ${\cal R} \equiv Y^2D^2/(\tr^2 \tY A_{2,Y}) \sim \pr^{4\lm_2 + 4 - 2\lm_3} / \tr^2 \sim \tr^p$ where $p > 0$ and $2\lm_2 + 2 > \lm_3$.
Substituting the solution near $\tr = 0$ for this case from eq.~(\ref{u1s}) into ${\cal R}$,
we get  ${\cal R} \ll 1$.
However, this solution also makes ${\cal R} < 1$  for a range $0 \leq \tr \lesssim 1$,
so that the approximated form of $\tilde\mu$ given in eq.~(\ref{mu1}) is valid for this range of $\tr$.
Hence, the solution in eq.~(\ref{u1s}) satisfies eq.~(\ref{boxph0}) for the case ${\cal R} \sim \tr^p$ as long as $0 \leq \tr \lesssim 1$, i.e., for $r \lesssim H_0^{-1}$.
In the opposite case where $1 / {\cal R} \sim \tr^{p}$ and $2\lm_2 + 2 < \lm_3$,
one can also check that the solution in eq.~(\ref{u2s}) satisfies eq.~(\ref{boxph0}) for a range $0 \leq \tr \lesssim 1$.
It is straightforward to show that, if ${\cal R} \sim$ constant, eq.~(\ref{boxph0}) gives $\pr \sim \tr^{1/(2\lm_3 - 1)}$ with $3\lm_3 = 2\lm_2 + 3$.
Moreover, $\pr \sim \tr^p$ where $p\geq 0$ can also satisfy eq.~(\ref{boxph0}) for the case $A_2 = A_K(Y)A_V(\phi)$ when $0\leq \tr \lesssim 1$.
According to these analysis, we conclude that, for a given value of parameters $\lm_1, \cdots, \lm_4$,
$\pr$ obeys the same relation $\pr \sim \tr^p$ for all distance inside the Huble radius.
This implies that the Vainshtein mechanism disappears in disformal gravity.

To determine  deviation of the ratio $-\Phi/\Psi$ from unity,
we first estimate the magnitude of $\Phi$ and $\Psi$ in terms of our variables.
Dividing eqs.~(\ref{eq:00w}) and (\ref{eq:11w}) by $H_0^2$ and ignoring the contributions from scalar field for a while,
the resulting equations give
\be
\Phi \simeq - \Psi \simeq \frac 16 \tr_s^2\trho_s \frac{\tr_s}{\tr} \sim 10^{-6} \frac{\tr_s}{\tr}\,,
\label{phipsi}
\ee
where $\tr \geq \tr_s$ and $\trho_s$ is the mean density of the sun divided by $H_0^2$.
Hence, in the vicinity of the sun, $\Phi \sim - \Psi \sim 10^{-6}$.
Using the solutions in eqs.~(\ref{u1s}) and (\ref{u2s}),
it can be shown that $\tY^{\lm_3} \sim \tr^{q_1}$ and $|DY| \sim \tr^{q_2}$ where $q_1, q_2 >1$,
so that these quantities are less than $10^{-14}$ inside solar system.
From eqs.~(\ref{u1s}) and (\ref{u2s}), we respectively get $\tr\pr \sim \tY^{\lm_3}$ and $\tr \pr \sim |D Y|$.
Based on these results and the fact that $\pr \prr \sim \tY /\tr$,
the third term in eq.~(\ref{eq:00w}) is negligible compared with the fourth term inside the solar system.
Furthermore, using $\tY^{\lm_3} \sim \tr^{q_1}$ and $\e^{-\lm_4 \phi} \lesssim 1$,
$A_2$ and also $Y A_{2,Y}$ in eqs.~(\ref{eq:00w}) and (\ref{eq:11w}) can be neglected.
According to this consideration,
we conclude that the contributions from scalar field in eqs.~(\ref{eq:00w}) and (\ref{eq:11w}) are negligible inside the solar system.
We also obtain the same conclusion when we consider the distances that are larger or comparable with the size of Milky Way.
Hence, the Vainshtein mechanism is not necessary for disformal gravity.
This conclusion is in agreement with \cite{Sakstein:14isa,Sakstein:152}.

\subsection{Cosmological background}

In order to study the screening mechanism in the cosmological, i.e., FLRW, background,
we assume that the perturbations around FLRW background have spherical symmetry and write the line element as
\ba
ds^2 &=& - \(1 + 2 \Psi(t,r)\) dt^2
\label{dstime}\\
&&+ a^2\[\(1 + 2\Phi(t,r)\)\,dr^2 +r^2 \(d\theta^{2}+\sin^{2}\theta\,d\varphi^{2}\)\]\,,
\nonumber
\ea
where $\Phi$ and $\Psi$ now are the metric perturbations.
Using this line element and decomposing the field $\phi$ and energy density of matter $\rho_m$ into background and perturbed parts as
\be
\phi\rightarrow \phi (t)+\pi(t, r)\,,
\quad
\rho\rightarrow \rho (t)\left[ 1+\delta (t,r)\right]\,,
\ee
the effective action for perturbations for the theory described by action (\ref{act-glpv}) can be constructed.
Based on discussions in \cite{Koba},
the non-linear perturbations on small scales obey the following relations:
\begin{eqnarray}
0&=&
2\xi_2 x_1 + {\cal G} x_2 - {\cal F} x_3 +\alpha_2 x_1^2 + 2\alpha_* x_1\left(rx_1'+x_1\right)
\nonumber\\
&&
-\frac{2}{\sqrt{\Lambda} a^3}\partial_t\left( a^3\xi_t x_1\right)\,,\label{eq1}
\\
0 &=&
{\cal G}x_3-\xi_1x_1-\alpha_1 x_1^2 - A\,,\label{eq2}
\\
0 &=&
\eta x_1 -2\xi_1 x_2 + 4\xi_2 x_3 +2\mu x_1^2+2\nu x_1^3-4\alpha_1x_1x_2
\label{eq3}\\
&&
+4\alpha_2x_1x_3
-4\alpha_*\left(r x_1 x_3' + 3 x_1 x_3\right) 
+ \frac{4\xi_t}{\sqrt{\Lambda} a^2}\partial_t\left( a^2x_3\right)\,, 
\nonumber\\
\end{eqnarray}
where the expressions for coefficients ${\cal G}, {\cal F}, \al_1, \al_2, \cdots, \xi_t$ in terms of $G_2, G_3, G_4$ and $F_4$ are given in \cite{Koba,Kimura:11},
$\Lambda \sim H_0^2$ is a mass scale,
and dimensionless dynamical variables for the perturbations $x_1, x_2, x_3$ and $A$ are defined as
\ba
x_1(t,r) &\equiv& \frac{1}{\Lambda}\frac{\pi'}{a^2r}\,,
\quad
x_2(t,r) \equiv \frac{1}{\Lambda}\frac{\Psi'}{a^2r}\,,
\nonumber\\
x_3(t,r) &\equiv& -\frac{1}{\Lambda}\frac{\Phi'}{a^2r}\,,
\quad
A(t,r) \equiv \frac{1}{\Lambda}\frac{M(t,r)}{8 \pi r^3}\,.
\ea
Here, the mass inside the sphere of radius $r$ is
\begin{eqnarray}
M(t,r) \equiv \int^r_0 4\pi \bar r^2\rho(t) \delta(t,\bar r)\, d\bar r\,.
\end{eqnarray}
Substituting $x_2$ and $x_3$ from eqs.~(\ref{eq1}) and (\ref{eq2}) into eq.~(\ref{eq3}),
and write the coefficients of the equations in terms of $A_2, C_3, G_4$ and $F_4$ for disformal gravity,
we get
\be
A_0
+2\left(A_1 - \frac{\kappa_1}{4 Y} \right) x_1
+\frac{\kappa_2}{\left(\gamma ^2 \gamma _1-1\right) Y \dot{\phi}} x_1^2 = 0\,,
\label{eq:x eq}
\ee
where
\ba
\dot\phi A_0 &\equiv& A \left(4 \gamma _2 H-\gamma ^2 \dot{\gamma }_1\right)
-\gamma ^2 \gamma _1^2 \left(5 A H+\dot{A}\right)
\nonumber\\
&&
+ \gamma_1\left(2 A H -A \dot{\gamma} \gamma + \gamma^2\left(3 A H+\dot{A}\right)\right)\,,
\label{defa0}
\\
A_1 &\equiv&
   -\frac{\gamma^2 \left(\gamma_1-1\right) \gamma_1 \Lambda\left(r A'+3 A\right)}{2 Y}
\label{defa1}
\ea
\ba
\kappa_1 &\equiv&
\gamma \Big[ 2 \gamma ^6 \gamma _1^6-\gamma ^4 \left(2 \gamma^2+3\right) \gamma_1^5
\nonumber\\
&&
+\gamma ^2 \gamma _1^4 \left(\gamma ^4+\gamma ^2\left(-8 \gamma _2+11 H^2+3\right)+4 \dot{\gamma } \gamma 
    H+1\right)
\nonumber\\
&&
-\gamma  \gamma _1^3 \Big(\gamma ^3 \left(-8 \gamma _2+7 H^2-4 \dot{\gamma }_1 H+1\right)
\nonumber\\
&&
+\gamma  \left(-6 \gamma _2+18 H^2+1\right)-6
    \dot{\gamma } \gamma ^2 H-2 \dot{\gamma } H\Big)
\nonumber\\
&&
+\gamma _1^2\Big(\gamma ^2 \left(8 \gamma _2^2-2 \gamma _2 \left(19 H^2+3\right)+2 H
    \left(H-4 \dot{\gamma }_2\right)\right)
    \nonumber\\
&&
    +8 H^2+\gamma ^4 \left(H \left(2
    \dot{\gamma }_1+H\right)-2 \gamma _2\right)-2 \dot{\gamma } \gamma ^3
    H\nonumber\\
&&+\gamma  H \left(2 \dot{\gamma } \left(4 \gamma _2-5\right)+3 H \Omega
    _m\right)\Big)+2 \gamma _1 \bigg(-4 \gamma _2^2 \gamma^2 \nonumber \\
&&-\left(\left(\gamma ^2+1\right) \dot{\gamma }_1-4 \dot{\gamma }_2\right)
    \gamma ^2 H+2 \gamma _2 H \Big(-6 \dot{\gamma } \gamma 
    \nonumber\\
&&
    +\gamma ^2 \left(2
    \dot{\gamma }_1+3 H\right)+8 H\Big)\bigg)+16 \gamma _2 H \left(2 \gamma
    _2 H-\gamma ^2 \dot{\gamma }_1\right)\Big]\,,\nonumber \\
\label{full-kp1}
\ea
\ba
   \kappa_2 &\equiv& \gamma^2 \gamma_1 \Lambda \left(2 \dot{\gamma} \left(\gamma_1 - 1\right) \gamma_1^2 \gamma^2-6 \dot{\gamma} \left(\gamma_1-1\right)\gamma_1
\right.
\nonumber\\
&&\left. +2 \gamma _1 \gamma ^3 \left(\gamma _1 \left(\dot{\gamma}_1 - \Dp Y \dot{\phi} - 3 \gamma_2 H+3 H\right)
\right.\right.
\\
&&\left.\left.
+ \Dp Y \dot{\phi}-3 \gamma_1^2 H+4 \gamma_2 H\right)
+3 \left(\gamma_1-1\right) \gamma_1^3 \gamma^5 H
\right.
\nonumber\\
&&\left.
+\gamma\left(3 \gamma_1^2 H+\gamma_1 \left(-2 \dot{\gamma}_1+6 \gamma_2 H-3 H\right)
-8 \gamma _2 H\right)\right)\,.\nonumber
\label{full-kp2}
\ea
In the above expressions,
all of the coefficients are evaluated using background quantities, e.g., $Y = - \dot\phi^2$,
and the dimensionless quantities $\gamma_1$ and $\gamma_2$ defined as
\be
\gamma_1 \equiv 1 - \DY Y^2\,,
\quad
\gamma_2 \equiv 2\DY Y^2 + \DYY Y^3\,.
\ee
Since $\ga^2 \lesssim 1$ through out the evolution of the universe and $D$ is supposed to be a polynomial function of $Y$,
we have $DY \sim \DY Y^2 \sim \DYY Y^3 < 1$.
Therefore, we can expand $\ga$ and $\ga_1$ around unity,
and consequently , up to leading order, eqs.~(\ref{defa0}) -- (\ref{full-kp2}) become
\ba
A_0 &\simeq& \frac{A \left(2 d_0+5 d_1+4 d_2\right) H+\dot{A} d_1}{\dot{\phi }}\,,
\nonumber\\
A_1 &\simeq&  \frac{d_1 \Lambda  \left(r A'+3 A\right)}{2 Y}\,,
\nonumber\\
- \frac{\kappa_1}{4 Y} &\simeq & \frac{3 H^2}{4 Y}\(1 - \Omega_m\)\,,
\nonumber\\
\frac{\kappa_2}{\left(\gamma ^2 \gamma _1-1\right) Y \dot{\phi}} &\simeq& 
\frac{2 \Lambda H d_2}{\dot\phi Y}\,,
\label{smallx}
\ea
where
\be
d_0 \equiv DY\,,
\quad
d_1 \equiv \DY Y^2\,,
\,\,\,\mbox{and}\,\,\,
d_2 \equiv \ga_2\,,
\ee
For illustration, we consider the form of $D$ given in eq.~(\ref{disf-mod}),
so that we get
\ba
A_0 &\sim& A \frac{H}{\dot\phi} Y_H^{1 + \lm_2}\,,
\nonumber\\
A_1 &\sim & \left(r A'+3 A\right) Y_H^{\lm_2}\,, 
\nonumber\\
- \frac{\kappa_1}{4 Y} &\sim & \(1 - \Omega_m\) \frac{H^2}{Y}\,,
\nonumber\\
\frac{\kappa_2}{\left(\gamma ^2 \gamma _1-1\right) Y \dot{\phi}} &\sim & 
\frac{H}{\dot\phi} Y_H^{\lm_2}\,,
\label{smallx1}
\ea
where $Y_H^\al$ denotes quantity whose magnitude is of order of $(Y / H_0^2)^\alpha$ and we have assumed that $\dot A \sim {\cal O}\(H A\)$.
Since $\phi$ always slowly evolves, one expects that $Y / H_0^2 \ll 1$ and $\dot\phi / H \ll 1$.
We first consider the case where $A \gg 1$ in which the Vainshtein mechanism works for general consideration in \cite{Koba}.
In the case where $|A_1| \gg |\kappa_1 / (4 Y)|$,
eq.~(\ref{eq:x eq}) yields,
\be
x_1 \sim {\cal O}\(A \frac{\dot\phi}{H}\)\,.
\ee
Using the approximation for the case of $\ga < 1$ as above,
we can write eqs.~(\ref{eq1}) and (\ref{eq2}) respectively as
\ba
x_2&\simeq&
x_3 
-x_1 \frac{H}{\dot\phi} Y_H^{\lm_2 +1}
- x_1^2 Y_H^{\lm_2}
\label{eq11}\\
&&
- x_1\left(rx_1'+x_1\right)Y_H^{\lm_2}
+ \dot{x}_1 Y_H^{\lm_2+ 1/2}\,,
\nonumber\\
x_3 &\simeq& A
+ x_1 \frac{H}{\dot\phi} Y_H^{\lm_2+1}
+x_1^2 Y_H^{\lm_2}\,.
\label{eq21}
\ea
The above equations suggest that the disformal gravity mimics the Einstein theory up to the small factors that are proportional to small ratio $Y / H_0^2$ without the help of non-linearity in $x_1$, i.e., without Vainshtein mechanism.
The non-linear term of $x_1$ in eq.~(\ref{eq21}) can give contribution to $x_3$ if $A$ is significantly large such that $A \gtrsim |Y_H^{- \lm_2} H^2 / Y|$.
In such case, the non-linearity in $x_1$ leads to large deviation from the Einstein theory instead of the Vainshtein mechanism.

Let us now consider the case where $|A_1| \ll |\kappa_1 / (4 Y)|$.
In this limit, eq.~(\ref{eq:x eq}) has two different solutions:
\be
x_1 \simeq  \left\{
\begin{array}{c}
x_1^{(1)} = A \frac{H / \dot\phi}{1 -  \Omega_m} Y_H^{\lm_2}
\\
x_1^{(2)} = 2\(1 - \Omega_m\) \frac{H^2}Y
\end{array}\right.\,.
\label{x1sol2}
\ee
It follows from the above equation that, for the first solution $x_1 = x_1^{(1)}$,
$x_1 \ll A$, while, for the second solution $x_1 = x_1^{(2)}$, $x_1 \gg 1$.
Similar to above discussion,
the disformal gravity mimics the Einstein theory when $x_1 \ll A$,
and the non-linearity in $x_1$ leads to large deviation from the Einstein  theory when $x_1 \gg 1$.

We conclude that the Vainshtein mechanism is absent in disformal gravity theory
although the action for this theory contains $L_3$.


\section{Conclusions}
\label{concl}

In this work, we study the Vainshtein mechanism and the evolution of background universe for a general purely disformal gravity theory in which the gravity action is a result from purely disformal transformation on the Einstein-Hilbert action.
We write the gravity action in the form of the GLPV theory and find that $A_3=0$ for this disformal gravity theory.
We discuss the cosmic evolution for this model of gravity,
and find that the accelerated expansion of the universe cannot be driven by kinetic terms of the scalar field as in the Galileon theory , i.e., self accelerating solution does not exist.
The accelerated expansion of the late-time universe can be achieved if the Lagrangian of the scalar field $A_2$ satisfies $A_2 / (2 X A_{2,X} - A_2) < - 1/3$.
The cosmic evolution for disformal gravity is nearly similar to that for $\Lambda$CDM model.

We then  study behavior of disformal gravity under spherically symmetric assumption for both static and FLRW background.
Based on evolution equations that have been derived in the literature,
we analyze their solutions for disformal gravity.
We have found that those solutions cannot provide Vainshtein mechanism although the action for disformal gravity contains the kinetic self interacting terms like $f(\phi,Y) \Box\phi$.
However, the absence of Vainshtein mechanism does not lead to serious problem in disformal gravity,
because, in the static spacetime, the disformal gravity nearly mimics the Einstein theory at all scales inside the Hubble radius without the help of Vainshtein mechanism.
This properti of disformal gravity is also hold for the cosmological background,
in which the small deviation from the Einstein theory arises from the slow evolution of the scalar field.
At leading order, this small deviation is negligible.

\begin{acknowledgements}
K.K. is supported by Thailand Research Fund (TRF) through grant RSA5780053.  DFM is supported by the Research Council of Norway.
\end{acknowledgements}

\appendix

\section{Riemann tensor}
\label{riemann-cal}

From eq.~(\ref{kamn}), we get
\ba
{}&&\mc K ^\al_{\ph \lm [\mu} \mc K^\lm _{\ph \nu]\bt}
=
\tfrac{1}{4} \gamma^2 \Bigl[\phi{}^{;\al} \phi{}_{;\bt} \bigl(- \phi{}_{;\nu} (D{}_{;\mu} \phi{}_{;\iota}+ 2 D \phi{}_{;\mu}{}_{;\iota})\nonumber\\
&&  + \phi{}_{;\mu} (D{}_{;\nu} \phi{}_{;\iota} + 2 D \phi{}_{;\nu}{}_{;\iota})\bigr) D{}^{;\iota} + D{}^{;\al} \bigl(\phi{}_{;\bt} (- \phi{}_{;\mu} D{}_{;\nu}  
\nonumber \\
&&+ D{}_{;\mu} \phi{}_{;\nu})
+ 2 D (\phi{}_{;\bt}{}_{;\mu} \phi{}_{;\nu} -  \phi{}_{;\mu} \phi{}_{;\bt}{}_{;\nu})\bigr) \phi{}_{;\iota} \phi{}^{;\iota}\Bigr]\nonumber\\
&&+ \tfrac{1}{4} \gamma^4 \phi{}^{;\al} \Bigl[2 D \phi{}_{;\mu} \phi{}_{;\bt}{}_{;\nu} \phi{}_{;\lm} D{}^{;\lm}
-  D{}_{;\bt} \phi{}_{;\mu} D{}_{;\nu} \phi{}_{;\lm} \phi{}^{;\lm}\nonumber\\
&&
  + D{}_{;\bt} D{}_{;\mu} \phi{}_{;\nu} \phi{}_{;\lm} \phi{}^{;\lm} 
+ 2 D D{}_{;\mu} \phi{}_{;\bt}{}_{;\nu} \phi{}_{;\lm} \phi{}^{;\lm} 
\nonumber\\
&&
+ 2 D D{}_{;\bt} \phi{}_{;\nu} \phi{}_{;\mu}{}_{;\lm} \phi{}^{;\lm} + 4 D^2 \phi{}_{;\bt}{}_{;\nu} \phi{}_{;\mu}{}_{;\lm} \phi{}^{;\lm} \nonumber\\
&&
- 2 D D{}_{;\bt} \phi{}_{;\mu} \phi{}_{;\nu}{}_{;\lm} \phi{}^{;\lm} 
+ 2 D^2 \phi{}_{;\mu} \phi{}_{;\bt}{}_{;\nu} \phi{}_{;\lm} D{}^{;\lm} \phi{}_{;\iota} \phi{}^{;\iota} 
\nonumber\\
&&
- 2 D \phi{}_{;\bt}{}_{;\mu} \bigl((D{}_{;\nu} \phi{}_{;\lm} 
+ 2 D \phi{}_{;\nu}{}_{;\lm}) \phi{}^{;\lm} + \phi{}_{;\nu} \phi{}_{;\lm} D{}^{;\lm} (1 + D \phi{}_{;\iota}\nonumber\\
&& \phi{}^{;\iota})\bigr) + \phi{}_{;\bt} \bigl(- D{}_{;\mu} (\phi{}_{;\nu} \phi{}_{;\lm} D{}^{;\lm}+ 2 D \phi{}_{;\nu}{}_{;\lm} \phi{}^{;\lm}) 
\nonumber\\
&&
+ 2 D (D{}_{;\nu} \phi{}_{;\mu}{}_{;\lm} \phi{}^{;\lm} + D \phi{}_{;\nu} \phi{}_{;\lm} D{}^{;\lm} \phi{}_{;\mu}{}_{;\iota} \phi{}^{;\iota}) 
\nonumber\\
&&+ \phi{}_{;\mu} \phi{}_{;\lm} D{}^{;\lm} (D{}_{;\nu} 
- 2 D^2 \phi{}_{;\nu}{}_{;\iota} \phi{}^{;\iota})\bigr)\Bigr]\,,
\label{one}
\ea
and get 
\ba
&&{}\nabla_{[\mu}\mc K^{\al}_{\ph \nu]\bt} =\tfrac{1}{2} \bigl(\phi{}_{;\bt} (- D{}^{;\al}{}_{;\mu} \phi{}_{;\nu} + \phi{}_{;\mu} D{}^{;\al}{}_{;\nu}) \nonumber\\
&&+ D{}^{;\al} (- \phi{}_{;\bt}{}_{;\mu} \phi{}_{;\nu} + \phi{}_{;\mu} \phi{}_{;\bt}{}_{;\nu})\bigr)
+\tfrac{1}{2} \gamma^2 \Bigl[D{}_{;\bt} \phi{}^{;\al}{}_{;\mu} \phi{}_{;\nu}
\nonumber\\
&&
 -  D{}_{;\bt} \phi{}_{;\mu} \phi{}^{;\al}{}_{;\nu} - 2 D \phi{}_{;\bt}{}_{;\mu} \phi{}^{;\al}{}_{;\nu} + 2 D \phi{}^{;\al}{}_{;\mu} \phi{}_{;\bt}{}_{;\nu} 
\nonumber\\
&&+ \phi{}_{;\bt} \bigl(- \phi{}^{;\al}{}_{;\nu} (D{}_{;\mu} + D \phi{}_{;\mu} \phi{}_{;\rho} D{}^{;\rho}) 
\nonumber\\
&&
+ \phi{}^{;\al}{}_{;\mu} (D{}_{;\nu} + D \phi{}_{;\nu} \phi{}_{;\rho} D{}^{;\rho})\bigr) 
+ \phi{}^{;\al} \bigl(2 D \phi{}_{;\bt}{}_{;\nu}{}_{;\mu} \nonumber\\
&&+ D{}_{;\bt}{}_{;\mu} \phi{}_{;\nu} -  \phi{}_{;\mu} D{}_{;\bt}{}_{;\nu} + D{}_{;\mu} \phi{}_{;\bt}{}_{;\nu}
\nonumber\\
&&
 - 2 D \phi{}_{;\bt}{}_{;\mu}{}_{;\nu} 
-  \phi{}_{;\bt} \phi{}_{;\mu} D{}_{;\nu} \phi{}_{;\rho} D{}^{;\rho} 
+ \phi{}_{;\bt} D{}_{;\mu} \phi{}_{;\nu} \phi{}_{;\rho} D{}^{;\rho} \nonumber\\
&&-  D \phi{}_{;\mu} \phi{}_{;\bt}{}_{;\nu} \phi{}_{;\rho} D{}^{;\rho} + D \phi{}_{;\bt} \phi{}_{;\nu} \phi{}_{;\mu}{}_{;\rho} D{}^{;\rho} 
\nonumber\\
&&
-  D \phi{}_{;\bt} \phi{}_{;\mu} \phi{}_{;\nu}{}_{;\rho} D{}^{;\rho} + \phi{}_{;\bt}{}_{;\mu} (- D{}_{;\nu} + D \phi{}_{;\nu} \phi{}_{;\rho} D{}^{;\rho})
\nonumber\\
&&
+ D \phi{}_{;\bt} \phi{}_{;\nu} D{}_{;\mu}{}_{;\rho} \phi{}^{;\rho} -  D \phi{}_{;\bt} \phi{}_{;\mu} D{}_{;\nu}{}_{;\rho} \phi{}^{;\rho}\bigr)\Bigr]
\nonumber\\
&&
+ \tfrac{1}{2} \gamma^4 \phi{}^{;\al} \biggl[D{}_{;\bt} \Bigl(- \phi{}_{;\nu} \bigl(D{}_{;\mu} \phi{}_{;\rho}+ 2 D \phi{}_{;\mu}{}_{;\rho}\bigr) + \phi{}_{;\mu} \bigl(D{}_{;\nu} \phi{}_{;\rho}\nonumber\\
&&  + 2 D \phi{}_{;\nu}{}_{;\rho}\bigr)\Bigr) \phi{}^{;\rho} 
+ D \Bigl(-2 \phi{}_{;\bt} D{}_{;\nu} \phi{}_{;\mu}{}_{;\rho} \phi{}^{;\rho}\nonumber\\
&&
 - 4 D \phi{}_{;\bt}{}_{;\nu} \phi{}_{;\mu}{}_{;\rho} \phi{}^{;\rho}+ 2 \phi{}_{;\bt}{}_{;\mu} \bigl(D{}_{;\nu} \phi{}_{;\rho} \nonumber\\
&& + 2 D \phi{}_{;\nu}{}_{;\rho}\bigr) \phi{}^{;\rho} 
\nonumber\\
&&
+ \phi{}_{;\bt} \phi{}_{;\mu} D{}_{;\nu} \phi{}_{;\rho} D{}^{;\rho} \phi{}_{;\sigma} \phi{}^{;\sigma} - 2 D \phi{}_{;\bt} \phi{}_{;\nu} \phi{}_{;\rho} D{}^{;\rho} \phi{}_{;\mu}{}_{;\sigma} \phi{}^{;\sigma} 
\nonumber\\
&&
+ 2 D \phi{}_{;\bt} \phi{}_{;\mu} \phi{}_{;\rho} D{}^{;\rho} \phi{}_{;\nu}{}_{;\sigma} \phi{}^{;\sigma} 
-  D{}_{;\mu} \bigl(2 \phi{}_{;\bt}{}_{;\nu} \phi{}_{;\rho} \phi{}^{;\rho} 
\nonumber\\
&&+ \phi{}_{;\bt} (-2 \phi{}_{;\nu}{}_{;\rho} \phi{}^{;\rho} 
+ \phi{}_{;\nu} \phi{}_{;\rho} D{}^{;\rho} \phi{}_{;\sigma} \phi{}^{;\sigma})\bigr)\Bigr)\biggr]\,.
\label{two}
\ea
We now use eqs.~(\ref{one}) and (\ref{two}) to compute Riemann tensor as
\ba
&&{} \bar R^\al_{\phantom{\al}\bt\mu\nu}
=  R^\al_{\ph\bt\mu\nu}
 + \nabla_{[\mu}\mc K^{\al}_{\ph \nu]\bt}
+ \mc K ^\al_{\ph \gamma [\mu} \mc K^\gamma _{\ph \nu]\bt} 
\nonumber\\
&&= R^\al_{\ph\bt\mu\nu}
+ \tfrac{1}{2} \bigl[\phi{}_{;\bt} (- D{}^{;\al}{}_{;\mu} \phi{}_{;\nu} + \phi{}_{;\mu} D{}^{;\al}{}_{;\nu}) \nonumber\\
&&+ D{}^{;\al} (- \phi{}_{;\bt}{}_{;\mu} \phi{}_{;\nu} + \phi{}_{;\mu} \phi{}_{;\bt}{}_{;\nu})\bigr]
+\tfrac{1}{4} \gamma^2 \biggl[\phi{}^{;\al} \Bigl(4 D \phi{}_{;\bt}{}_{;\nu}{}_{;\mu} \nonumber\\
&&
+ 2 D{}_{;\bt}{}_{;\mu} \phi{}_{;\nu} - 2 \phi{}_{;\mu} D{}_{;\bt}{}_{;\nu} + 2 D{}_{;\mu} \phi{}_{;\bt}{}_{;\nu} - 4 D \phi{}_{;\bt}{}_{;\mu}{}_{;\nu} \nonumber\\
&&-  \phi{}_{;\bt} \phi{}_{;\mu} D{}_{;\nu} \phi{}_{;\rho} D{}^{;\rho}
 + \phi{}_{;\bt} D{}_{;\mu} \phi{}_{;\nu} \phi{}_{;\rho} D{}^{;\rho} 
\nonumber\\
&&
- 2 D \phi{}_{;\mu} \phi{}_{;\bt}{}_{;\nu} \phi{}_{;\rho} D{}^{;\rho} - 2 \phi{}_{;\bt}{}_{;\mu} \bigl(D{}_{;\nu} -  D \phi{}_{;\nu} \phi{}_{;\rho} D{}^{;\rho}\bigr) 
\nonumber\\
&&
+ 2 D \phi{}_{;\bt} \phi{}_{;\nu} D{}_{;\mu}{}_{;\rho} \phi{}^{;\rho} - 2 D \phi{}_{;\bt} \phi{}_{;\mu} D{}_{;\nu}{}_{;\rho} \phi{}^{;\rho}\Bigr) \nonumber\\
&&+ \phi{}_{;\bt} \Bigl(2 \phi{}^{;\al}{}_{;\mu} \bigl(D{}_{;\nu} + D \phi{}_{;\nu} \phi{}_{;\rho} D{}^{;\rho}\bigr) -  \phi{}_{;\mu} \phi{}_{;\rho} \bigl(2 D \phi{}^{;\al}{}_{;\nu} D{}^{;\rho}
\nonumber\\
&&
 + D{}^{;\al} D{}_{;\nu} \phi{}^{;\rho}\bigr) + D{}_{;\mu} \bigl(-2 \phi{}^{;\al}{}_{;\nu} + D{}^{;\al} \phi{}_{;\nu} \phi{}_{;\rho} \phi{}^{;\rho}\bigr)\Bigr)
\nonumber\\
&&
 + 2 \Bigl(D{}_{;\bt} \bigl(\phi{}^{;\al}{}_{;\mu} \phi{}_{;\nu} -  \phi{}_{;\mu} \phi{}^{;\al}{}_{;\nu}\bigr)
+ D \bigl(\phi{}_{;\bt}{}_{;\nu} (2 \phi{}^{;\al}{}_{;\mu}\nonumber\\
&&
  -  D{}^{;\al} \phi{}_{;\mu} \phi{}_{;\rho} \phi{}^{;\rho}) + \phi{}_{;\bt}{}_{;\mu} (-2 \phi{}^{;\al}{}_{;\nu} + D{}^{;\al} \phi{}_{;\nu} \phi{}_{;\rho} \phi{}^{;\rho})\bigr)\Bigr)\biggr]
\nonumber\\
&&
+ \tfrac{1}{4} \gamma^4 \phi{}^{;\al} \biggl[2 D \phi{}_{;\mu} \phi{}_{;\bt}{}_{;\nu} \phi{}_{;\rho} D{}^{;\rho} + D{}_{;\bt} \phi{}_{;\mu} D{}_{;\nu} \phi{}_{;\rho} \phi{}^{;\rho} 
\nonumber\\
&&
-  D{}_{;\bt} D{}_{;\mu} \phi{}_{;\nu} \phi{}_{;\rho} \phi{}^{;\rho} - 2 D D{}_{;\mu} \phi{}_{;\bt}{}_{;\nu} \phi{}_{;\rho} \phi{}^{;\rho} - 2 D D{}_{;\bt} \phi{}_{;\nu} \phi{}_{;\mu}{}_{;\rho} \phi{}^{;\rho} 
\nonumber\\
&&
- 4 D^2 \phi{}_{;\bt}{}_{;\nu} \phi{}_{;\mu}{}_{;\rho} \phi{}^{;\rho} + 2 D D{}_{;\bt} \phi{}_{;\mu} \phi{}_{;\nu}{}_{;\rho} \phi{}^{;\rho} + 2 D^2 \phi{}_{;\mu} \phi{}_{;\bt}{}_{;\nu} \phi{}_{;\rho} 
\nonumber\\
&&\times D{}^{;\rho} \phi{}_{;\iota} \phi{}^{;\iota}
 - 2 D \phi{}_{;\bt}{}_{;\mu} \Bigl(- \bigl(D{}_{;\nu} \phi{}_{;\rho} + 2 D \phi{}_{;\nu}{}_{;\rho}\bigr) \phi{}^{;\rho} 
\nonumber\\
&&
+ \phi{}_{;\nu} \phi{}_{;\rho} D{}^{;\rho} \bigl(1 
+ D \phi{}_{;\iota} \phi{}^{;\iota}\bigr)\Bigr) + \phi{}_{;\bt} \Bigl(-2 D \bigl(D{}_{;\nu} \phi{}_{;\mu}{}_{;\rho} \phi{}^{;\rho} 
\nonumber\\
&&
+ D \phi{}_{;\nu} \phi{}_{;\rho} D{}^{;\rho} \phi{}_{;\mu}{}_{;\iota} \phi{}^{;\iota}\bigr) + \phi{}_{;\mu} \phi{}_{;\rho} D{}^{;\rho} \bigl(2 D^2 \phi{}_{;\nu}{}_{;\iota} \phi{}^{;\iota} + D{}_{;\nu} (1 +
\nonumber\\
&&
+ 2 D \phi{}_{;\iota} \phi{}^{;\iota})\bigr) -  D{}_{;\mu} \bigl(-2 D \phi{}_{;\nu}{}_{;\rho} \phi{}^{;\rho} + \phi{}_{;\nu} \phi{}_{;\rho} D{}^{;\rho} (1+
\nonumber\\
{}&& + 2 D \phi{}_{;\iota} \phi{}^{;\iota})\bigr)\Bigr)\biggr]\,.
\label{riemann}
\ea
Inserting the expression for the barred Riemann tensor $\bar{R}^\al_{\phantom{\al}\bt\al\nu}$ from the above equation into eq.~(\ref{ricci-form}), we get
\ba
{}&& \bar{R}=R - \ga^2\phi^{;\al}\phi^{;\bt} R_{\al\bt}+\tfrac{1}{2} \bigl[2 X D{}^{;\al}{}_{;\al} + \phi{}_{;\al} D{}^{;\al} \phi{}^{;\nu5}{}_{;\nu5} +
\nonumber\\
&&
 (\phi{}^{;\al} D{}_{;\al}{}_{;\nu5} -  D{}^{;\al} \phi{}_{;\al}{}_{;\nu5}) \phi{}^{;\nu5}\bigr]
+\tfrac{1}{2} \gamma^2 \bigl[2 X (1 
+ 2 D X) D{}^{;\al}{}_{;\al} 
\nonumber\\
&&
+ 2 X^2 D{}_{;\al} D{}^{;\al} 
+ 2 D \phi{}^{;\rho}{}_{;\rho}{}_{;\al} \phi{}^{;\al}
+ 2 D \phi{}^{;\al}{}_{;\al} \phi{}^{;\rho}{}_{;\rho} + 3 \phi{}_{;\al} D{}^{;\al} \phi{}^{;\rho}{}_{;\rho}
\nonumber\\
&&
 + 2 D X \phi{}_{;\al} D{}^{;\al} \phi{}^{;\rho}{}_{;\rho} - 2 D \phi{}^{;\al} \phi{}_{;\al}{}^{;\rho}{}_{;\rho} + X \phi{}_{;\al} D{}^{;\al} \phi{}_{;\rho} D{}^{;\rho} 
\nonumber\\
&&
+ \phi{}^{;\al} D{}_{;\al}{}_{;\rho} \phi{}^{;\rho} + 2 D X \phi{}^{;\al} D{}_{;\al}{}_{;\rho} \phi{}^{;\rho} - 3 D{}^{;\al} \phi{}_{;\al}{}_{;\rho} \phi{}^{;\rho}  
\nonumber\\
&&- 4 D X D{}^{;\al} \phi{}_{;\al}{}_{;\rho} \phi{}^{;\rho}
- 2 D \phi{}_{;\al}{}_{;\rho} \phi{}^{;\al}{}^{;\rho} -  D \phi{}_{;\al} D{}^{;\al} \phi{}^{;\rho} \phi{}_{;\rho}{}_{;\iota} \phi{}^{;\iota}\bigr]
\nonumber\\
&&
+\tfrac{1}{2} \gamma^4 \Bigl[2 X^2 \bigl(1 + 2 D X\bigr) D{}_{;\al} D{}^{;\al} + \phi{}_{;\al} D{}^{;\al} \bigl(4 D X \phi{}^{;\rho}{}_{;\rho} 
\nonumber\\
&&+ (1 + 2 D X) (X \phi{}_{;\rho} D{}^{;\rho} 
-  D \phi{}^{;\rho} \phi{}_{;\rho}{}_{;\iota} \phi{}^{;\iota})\bigr) - 2 D \phi{}^{;\rho} \bigl(X (3  
\nonumber\\
&&+ 2 D X) D{}^{;\al} \phi{}_{;\al}{}_{;\rho} + 2 D \phi{}^{;\al} (\phi{}_{;\al}{}_{;\rho} \phi{}^{;\iota}{}_{;\iota}
\nonumber\\
&&
-  \phi{}_{;\rho}{}_{;\iota} \phi{}_{;\al}{}^{;\iota})\bigr)\Bigr]\,. 
\label{ricci1}
\ea
The terms that are proportional to $2XD$ in eq.~(\ref{ricci1}) can be simplified by writing $2XD = 1 - 1 / \ga^2$,
so that this equation becomes
\ba
{}&& \bar{R} = R - \ga^2\phi^{;\al}\phi^{;\bt} R_{\al\bt}
+ \tfrac{1}{2} D{}^{;\al} \phi{}_{;\al}{}_{;\rho} \phi{}^{;\rho}
+\gamma^2 \Bigl[2 X D{}^{;\al}{}_{;\al}\nonumber\\
&& + D \phi{}^{;\rho}{}_{;\rho}{}_{;\al} \phi{}^{;\al} + D \phi{}^{;\al}{}_{;\al} \phi{}^{;\rho}{}_{;\rho} + \phi{}_{;\al} D{}^{;\al} \phi{}^{;\rho}{}_{;\rho} -  D \phi{}^{;\al} \phi{}_{;\al}{}^{;\rho}{}_{;\rho} 
\nonumber\\
&&
+ \phi{}^{;\al} D{}_{;\al}{}_{;\rho} \phi{}^{;\rho} 
-  D{}^{;\al} \phi{}_{;\al}{}_{;\rho} \phi{}^{;\rho} -  D \phi{}_{;\al}{}_{;\rho} \phi{}^{;\al}{}^{;\rho}\Bigr]
\nonumber\\
&&
+ \tfrac{1}{2} \gamma^4 \Bigl[4 X^2 D{}_{;\al} D{}^{;\al} + 2 \phi{}_{;\al} D{}^{;\al} \bigl(\phi{}^{;\rho}{}_{;\rho} + X \phi{}_{;\rho} D{}^{;\rho} 
\nonumber\\
&&-  D \phi{}^{;\rho} \phi{}_{;\rho}{}_{;\iota} \phi{}^{;\iota}\bigr) 
-  \phi{}^{;\rho} \bigl((3 + 4 D^2 X^2) D{}^{;\al} \phi{}_{;\al}{}_{;\rho} \nonumber\\
&&+ 4 D^2 \phi{}^{;\al} (\phi{}_{;\al}{}_{;\rho} \phi{}^{;\iota}{}_{;\iota} -  \phi{}_{;\rho}{}_{;\iota} \phi{}_{;\al}{}^{;\iota})\bigr)\Bigr]\,.
\label{ricci-sim1}
\ea

\section{Integration by parts}
\label{byparts}

We can simplify the action in eq.~(\ref{act0})
by performing the following integration by parts:
\ba
S &=& \frac{M_p^2}{2}\int d^4 x \sqrt{-g}\biggl\{
\frac 1{\ga} R - \ga\phi^{;\al}\phi^{;\bt} R_{\al\bt}
+ \tfrac{1}{2\ga} D{}^{;\al} \phi{}_{;\al}{}_{;\om} \phi{}^{;\om}
\nonumber\\
&&
+\gamma \Bigl[2 X D{}^{;\al}{}_{;\al} + D \phi{}^{;\om}{}_{;\om}{}_{;\al} \phi{}^{;\al} + D \phi{}^{;\al}{}_{;\al} \phi{}^{;\om}{}_{;\om} \nonumber\\
&& + \phi{}_{;\al} D{}^{;\al} \phi{}^{;\om}{}_{;\om}-  D \phi{}^{;\al} \phi{}_{;\al}{}^{;\om}{}_{;\om} 
+ \phi{}^{;\al} D{}_{;\al}{}_{;\om} \phi{}^{;\om} \nonumber\\
&&
-  D{}^{;\al} \phi{}_{;\al}{}_{;\om} \phi{}^{;\om} -  D \phi{}_{;\al}{}_{;\om} \phi{}^{;\al}{}^{;\om}\Bigr]
+ \tfrac{1}{2} \gamma^3 \Bigl[
\underbrace{
4 X^2 D{}_{;\al} D{}^{;\al} 
}_{A}\nonumber\\
&&
+ 2 \phi{}_{;\al} D{}^{;\al} \bigl(\phi{}^{;\om}{}_{;\om} + X \phi{}_{;\om} D{}^{;\om} -  D \phi{}^{;\om} \phi{}_{;\om}{}_{;\lm} \phi{}^{;\lm}\bigr) 
\nonumber\\
&&
-  \phi{}^{;\om} \bigl((3 + 4 D^2 X^2) D{}^{;\al} \phi{}_{;\al}{}_{;\om} + 4 D^2 \phi{}^{;\al} (\phi{}_{;\al}{}_{;\om} \phi{}^{;\lm}{}_{;\lm} \nonumber\\
&&-  \phi{}_{;\om}{}_{;\lm} \phi{}_{;\al}{}^{;\lm})\bigr)\Bigr]
\biggr\}\,.
\label{act1}
\ea
Since
\be
4\nabla_\al\(\frac{1}{\sqrt{1 - 2X D}}\) D^{;\al} X 
= \frac{4 X D{}^{;\al} (X D{}_{;\al} -  D \phi{}_{;\al}{}_{;\bt} \phi{}^{;\bt})}{(1 - 2 D X)^{\tfrac{3}{2}}} \,,
\ee
we can write term $A$ in the action (\ref{act1}) in terms of $\nabla_\al \frac{1}{\sqrt{1 - 2XD}}$ as,
and hence this action becomes
\ba
S &=& \frac{M_p^2}{2}\int d^4 x \sqrt{-g}\biggl\{
\frac R{\ga}  - \ga\phi^{;\al}\phi^{;\bt} R_{\al\bt}
+ \tfrac{1}{2\ga} D{}^{;\al} \phi{}_{;\al}{}_{;\om} \phi{}^{;\om}
\nonumber\\
&&
+\gamma \Bigl[2 X D{}^{;\al}{}_{;\al} + D \phi{}^{;\om}{}_{;\om}{}_{;\al} \phi{}^{;\al} + D \phi{}^{;\al}{}_{;\al} \phi{}^{;\om}{}_{;\om} 
\nonumber\\
&& + \phi{}_{;\al} D{}^{;\al} \phi{}^{;\om}{}_{;\om}-  D \phi{}^{;\al} \phi{}_{;\al}{}^{;\om}{}_{;\om} 
+ \phi{}^{;\al} D{}_{;\al}{}_{;\om} \phi{}^{;\om}
\nonumber\\
&& -  D{}^{;\al} \phi{}_{;\al}{}_{;\om} \phi{}^{;\om} -  D \phi{}_{;\al}{}_{;\om} \phi{}^{;\al}{}^{;\om}\Bigr]
\nonumber\\
&&\underbrace{
+\tfrac{1}{2} 4\nabla\_\al\(\frac{1}{\sqrt{1 - 2X D}}\) D^{;\al} X  
}_{A_1}
+ \frac 12 \gamma^3\biggl[2 \phi{}_{;\al} D{}^{;\al} \bigl(\phi{}^{;\om}{}_{;\om} \nonumber\\
&&+ X \phi{}_{;\om} D{}^{;\om} 
-  D \phi{}^{;\om} \phi{}_{;\om}{}_{;\lm} \phi{}^{;\lm}\bigr) 
\nonumber\\
&&
+ \phi{}^{;\om} \Bigl(\bigl(-2  \underbrace{
-\(1 - 2XD\)D{}^{;\al} \phi{}_{;\al}{}_{;\om}
}_{A_2}\nonumber\\
&&
+ 4 D^2 \phi{}^{;\al} (
- \phi{}_{;\al}{}_{;\om} \phi{}^{;\lm}{}_{;\lm} + \phi{}_{;\om}{}_{;\lm} \phi{}_{;\al}{}^{;\lm})\bigr)\Bigr)\biggr]\biggl\}\,.
\label{act11}
\ea
The term $A_2$ cancels with the term in the second line,
and the term $A_1$ in the above action can be integrated by parts,
so that this action becomes
\ba
S &=& \frac{M_p^2}{2}\int d^4 x \sqrt{-g}\biggl\{
\frac 1{\ga} R - \ga\phi^{;\al}\phi^{;\bt} R_{\al\bt}+ \gamma \Bigl[D \phi{}^{;\om}{}_{;\om}{}_{;\al} \phi{}^{;\al} 
\nonumber\\
&&
+ D \phi{}^{;\al}{}_{;\al} \phi{}^{;\om}{}_{;\om} + \phi{}_{;\al} D{}^{;\al} \phi{}^{;\om}{}_{;\om} -  D \phi{}^{;\al} \phi{}_{;\al}{}^{;\om}{}_{;\om} 
\nonumber\\
&&
+ \phi{}^{;\al} D{}_{;\al}{}_{;\om} \phi{}^{;\om} + D{}^{;\al} \phi{}_{;\al}{}_{;\om} \phi{}^{;\om} -  D \phi{}_{;\al}{}_{;\om} \phi{}^{;\al}{}^{;\om}\Bigr]
\nonumber\\
&&
+ \gamma^3 \Bigl[\phi{}_{;\al} D{}^{;\al} \bigl(\phi{}^{;\om}{}_{;\om} 
\underbrace{
+ X \phi{}_{;\om} D{}^{;\om}
}_{B}
 -  D \phi{}^{;\om} \phi{}_{;\om}{}_{;\lm} \phi{}^{;\lm}\bigr)
\nonumber\\
&&
 -  \phi{}^{;\om} \bigl(D{}^{;\al} \phi{}_{;\al}{}_{;\om} 
+ 2 D^2 \phi{}^{;\al} (\phi{}_{;\al}{}_{;\om} \phi{}^{;\lm}{}_{;\lm} -  \phi{}_{;\om}{}_{;\lm} \phi{}_{;\al}{}^{;\lm})\bigr)\Bigr]\,.\nonumber \\&&
\label{act2}
\ea
Since
\ba
&&\nabla_\al \(\frac{1}{\sqrt{1 - 2XD}}\)\phi^{;\al} D_{;\bt} \phi^{;\bt} \nonumber \\&&
=
- \frac{\phi{}_{;\al} D{}^{;\al} (- X \phi{}_{;\rho} D{}^{;\rho} + D \phi{}_{;\rho}{}_{;\bt} \phi{}^{;\bt} \phi{}^{;\rho})}{(1 - 2 D X)^{\tfrac{3}{2}}}\,, 
\ea
we can write the term $B$ in action (\ref{act2}) in terms of $\nabla_a \ga$,
so that this action becomes
\ba
S &=& \frac{M_p^2}{2}\int d^4 x \sqrt{-g}\biggl\{
\frac 1{\ga} R - \ga\phi^{;\al}\phi^{;\bt} R_{\al\bt}+ \gamma \Bigl[D \phi{}^{;\om}{}_{;\om}{}_{;\al} \phi{}^{;\al}
\nonumber\\
&&
 + D \phi{}^{;\al}{}_{;\al} \phi{}^{;\om}{}_{;\om} + \phi{}_{;\al} D{}^{;\al} \phi{}^{;\om}{}_{;\om} -  D \phi{}^{;\al} \phi{}_{;\al}{}^{;\om}{}_{;\om} 
\nonumber\\
&&
+ \phi{}^{;\al} D{}_{;\al}{}_{;\om} \phi{}^{;\om} + D{}^{;\al} \phi{}_{;\al}{}_{;\om} \phi{}^{;\om} -  D \phi{}_{;\al}{}_{;\om} \phi{}^{;\al}{}^{;\om}\Bigr]
\nonumber\\
&&
\underbrace{
+ \nabla_\al \(\frac{1}{\sqrt{1 - 2XD}}\)\phi^{;\al} D_{;\bt} \phi^{;\bt} 
}_{B_1}+ \gamma^3\Bigl[\phi{}_{;\al} D{}^{;\al} \phi{}^{;\om}{}_{;\om}\nonumber \\&&
 -  \phi{}^{;\om} \bigl(D{}^{;\al} \phi{}_{;\al}{}_{;\om} 
+ 2 D^2 \phi{}^{;\al} (\phi{}_{;\al}{}_{;\om} \phi{}^{;\lm}{}_{;\lm}\nonumber\\
&&
 -  \phi{}_{;\om}{}_{;\lm} \phi{}_{;\al}{}^{;\lm})\bigr)\Bigr]\,.
\label{act21}
\ea
The term $B_1$ can be integrated by parts,
so this action becomes
\ba
S &=& \frac{M_p^2}{2}\int d^4 x \sqrt{-g}\biggl\{
\frac 1{\ga} R - \ga\phi^{;\al}\phi^{;\bt} R_{\al\bt}
\nonumber\\
&&
+ D \gamma \(\phi{}^{;\om}{}_{;\om}{}_{;\al} \phi{}^{;\al} + \phi{}^{;\al}{}_{;\al} \phi{}^{;\om}{}_{;\om} -  \phi{}^{;\al} \phi{}_{;\al}{}^{;\om}{}_{;\om} -  \phi{}_{;\al}{}_{;\om} \phi{}^{;\al}{}^{;\om}\)
\nnb\\
&&
+ \gamma^3\Bigl[\phi{}_{;\al} D{}^{;\al} \phi{}^{;\om}{}_{;\om} -  \phi{}^{;\om} \bigl(D{}^{;\al} \phi{}_{;\al}{}_{;\om} 
\underbrace{
+ 2 D^2 \phi{}^{;\al} (\phi{}_{;\al}{}_{;\om} \phi{}^{;\lm}{}_{;\lm}
}_{C}\nonumber \\&&
 -  \phi{}_{;\om}{}_{;\lm} \phi{}_{;\al}{}^{;\lm})\bigr)\Bigr]\,.
\label{act3}
\ea
Since
\ba
&&2 \nabla_\al\(\frac{1}{\sqrt{1 - 2XD}}\) D \phi^{;\al} \Box \phi \nonumber \\&&
=
- \frac{D \phi{}^{;\bt}{}_{;\bt} \bigl(\phi{}_{;\al} D{}^{;\al} \phi{}_{;\lm} + 2 D \phi{}^{;\al} \phi{}_{;\al}{}_{;\lm}\bigr) \phi{}^{;\lm}}{\bigl(1 + D (\phi{}_{;\al} \phi{}^{;\al})\bigr)^{\tfrac{3}{2}}}\,,
\ea
we can write the term $C$ in action (\ref{act3}) in terms of $\nabla_a \ga$,
so that this action becomes
\ba
S &=& \frac{M_p^2}{2}\int d^4 x \sqrt{-g}\biggl\{
\frac 1{\ga} R - \ga\phi^{;\al}\phi^{;\bt} R_{\al\bt}+ D \gamma \Big(\phi{}^{;\om}{}_{;\om}{}_{;\al} \phi{}^{;\al}
\nonumber\\
&&
 + \phi{}^{;\al}{}_{;\al} \phi{}^{;\om}{}_{;\om} -  \phi{}^{;\al} \phi{}_{;\al}{}^{;\om}{}_{;\om} -  \phi{}_{;\al}{}_{;\om} \phi{}^{;\al}{}^{;\om}\Big)
\nonumber\\
&&
\underbrace{
+ 2 \nabla_\al\(\frac{1}{\sqrt{1 - 2XD}}\) D \phi^{;\al} \Box \phi 
}_{C_1}\nonumber \\&&
+ \gamma^3\bigl[\phi{}_{;\al} D{}^{;\al} (\phi{}^{;\om}{}_{;\om} + D \phi{}_{;\om} \phi{}^{;\om} \phi{}^{;\lm}{}_{;\lm}) 
\nonumber\\
&&
+ \phi{}^{;\om} (- D{}^{;\al} \phi{}_{;\al}{}_{;\om} 
\underbrace{
+ 2 D^2 \phi{}^{;\al} \phi{}_{;\om}{}_{;\lm} \phi{}_{;\al}{}^{;\lm}
}_{C_2}
)\bigr]\,.
\label{act31}
\ea
We now use
\ba
 &&-2\nabla_\al\(\frac{1}{\sqrt{1 - 2XD}}\) D \phi_{;\sig}{}^{;\al}\phi^{;\sig}\nonumber \\&&
=
\frac{D \bigl(2 D \phi{}^{;\al} \phi{}_{;\sig}{}_{;\bt} \phi{}_{;\al}{}^{;\bt} + D{}^{;\al} \phi{}_{;\al}{}_{;\bt} \phi{}^{;\bt} \phi{}_{;\sig}\bigr) \phi{}^{;\sig}}{\bigl(1 + D (\phi{}_{;\al} \phi{}^{;\al})\bigr)^{\tfrac{3}{2}}}\,,
\ea
to rewrite term $C_2$ in action (\ref{act31}),
such that the  action becomes
\ba
S &=& \frac{M_p^2}{2}\int d^4 x \sqrt{-g}\biggl\{
\frac R{\ga}  - \ga\phi^{;\al}\phi^{;\bt} R_{\al\bt}
+ D \gamma \Big(\phi{}^{;\om}{}_{;\om}{}_{;\al} \phi{}^{;\al}\nonumber\\
&&
 + \phi{}^{;\al}{}_{;\al} \phi{}^{;\om}{}_{;\om} -  \phi{}^{;\al} \phi{}_{;\al}{}^{;\om}{}_{;\om} -  \phi{}_{;\al}{}_{;\om} \phi{}^{;\al}{}^{;\om}\Big)
\nonumber\\
&&
+ \gamma D{}^{;\al} \bigl[\phi{}_{;\al} \phi{}^{;\om}{}_{;\om} - \phi{}^{;\om} \phi{}_{;\al}{}_{;\om}\bigr]
\label{act33}\\
&&
\underbrace{
+ 2 \nabla_\al\(\frac{1}{\sqrt{1 - 2XD}}\) D \phi^{;\al} \Box \phi 
}_{C_1}\nonumber \\&&
\underbrace{
- 2 \nabla_\al\(\frac{1}{\sqrt{1 - 2XD}}\) D \phi_{;\sig}{}^{;\al}\phi^{;\sig}
}_{C_3}
\,.
\nonumber
\ea
Terms $C_1$ and $C_3$ can be integrated by parts,
so that the above action becomes
\ba
&&S = \frac{M_p^2}{2}\int d^4 x \sqrt{-g}\biggl\{
\frac R{\ga}- \ga\phi^{;\al}\phi^{;\bt} R_{\al\bt}
+ \gamma D \times \nonumber\\
&&\Bigl[- \phi{}^{;\om}{}_{;\om}{}_{;\al} \phi{}^{;\al} 
- \phi{}^{;\al}{}_{;\al} \phi{}^{;\om}{}_{;\om} 
+ \phi{}^{;\al} \phi{}_{;\al}{}^{;\om}{}_{;\om} + \phi{}_{;\al}{}_{;\om} \phi{}^{;\al}{}^{;\om}\Bigr]\nonumber \\&&
+\ga\Bigl[
D{}^{;\al} \phi{}_{;\al}{}_{;\om} \phi{}^{;\om}- \phi{}_{;\al} D{}^{;\al} \phi{}^{;\om}{}_{;\om} \Bigr]\biggl\}\,.
\label{act34}
\ea
Using
\be
R_{\om\al} \phi^{;\om}\phi^{;\al}
=
\phi{}^{;\al} \phi{}_{;\al}{}^{;\om}{}_{;\om} 
- \phi{}^{;\om}{}_{;\om}{}_{;\al} \phi{}^{;\al} \,.
\ee
Action (\ref{act34}) becomes
\ba
S &=& \frac{M_p^2}{2}\int d^4 x \sqrt{-g}\biggl\{\,\,
\frac 1{\ga} R 
- \gamma D \Bigl[(
\(\Box \phi\)^2
- \phi{}_{;\al}{}_{;\bt} \phi{}^{;\al}{}^{;\bt}\Bigr]
\nonumber\\
&&+\ga\Bigl[
D{}^{;\al} \phi{}_{;\al}{}_{;\om} \phi{}^{;\om}
-
\phi{}_{;\al} D{}^{;\al} \phi{}^{;\om}{}_{;\om} 
\Bigr]\biggl\}\,.
\label{act35-a}
\ea

\section{the action in GLPV form}
\label{glpv}

Using the definition of $Y$ in eq.~(\ref{def-y}),
we can write $\gamma$ defined in eq.~(\ref{inverse}) as $\ga = 1 / \sqrt{1 + Y D(\phi,Y)}$.
Thus, if we set $G_4(\phi, Y) \equiv M_p^2 / (2 \gamma)$,
we can write the first line of eq.~(\ref{act35}) in the Horndeski form and the action becomes
\ba
S_g &=& \int d^4 x \sqrt{-g}\biggl\{\,\,
G_4 R 
- 2 G_{4,Y} \Bigl[
\(\Box \phi\)^2
- \phi{}_{;\al}{}_{;\bt} \phi{}^{;\al}{}^{;\bt}\Bigr]
\nonumber\\
&& \clipbox{-2 0 10 0}{$
\underbrace{
+\frac{M_p^2\ga}{2}\Bigl[
Y \DY\(
\(\Box \phi\)^2
- \phi{}_{;\al}{}_{;\bt} \phi{}^{;\al}{}^{;\bt}\)
+D{}^{;\al} \phi{}_{;\al}{}_{;\om} \phi{}^{;\om}\quad}_{e_1}$} \nnb \\&&
\clipbox{120 0 -2 0}{$\underbrace{\,\hphantom{\hspace{12em}}
-\phi{}_{;\al} D{}^{;\al} \phi{}^{;\om}{}_{;\om} 
\Bigr]}$}
\biggr\}\,.
\label{act-h}
\ea
Using $D^{;\al} = \Dp \phi^{;\al} + \DY Y^{;\al}$,
one can write the second line of eq.~(\ref{act-h}) as
\ba
e_1 &=&
\frac{M_p^2\ga}{2}\Bigl[
\DY\Big(
Y\big(\( \Box \phi \)^2
- \phi{}_{;\al}{}_{;\bt} \phi{}^{;\al}{}^{;\bt}\big)
+ Y^{;\al} \phi{}_{;\al}{}_{;\om} \phi{}^{;\om}
\nonumber\\
&&-
Y^{;\al}\phi{}_{;\al} \phi{}^{;\om}{}_{;\om} 
\Big)
+ \Dp \( \phi^{;\al} \phi{}_{;\al}{}_{;\om} \phi{}^{;\om}
-
\phi^{;\al}\phi{}_{;\al} \phi{}^{;\om}{}_{;\om}\)
\Bigr]
\nonumber\\
&=&
\frac{M_p^2\ga}{2}\Bigl[
\DY\Big(
Y\big(\(\Box \phi\)^2
- \phi{}_{;\al}{}_{;\bt} \phi{}^{;\al}{}^{;\bt}\big)
\nonumber\\
&&
+ 2\phi_{;\lm}\phi^{;\lm;\al} \phi{}_{;\al}{}_{;\om} \phi{}^{;\om}
- 2 \phi_{;\lm} \phi^{;\lm;\al}\phi{}_{;\al} \Box\phi
\Big)
\nonumber\\
&&
+ \Dp \( \phi^{;\al} \phi{}_{;\al}{}_{;\om} \phi{}^{;\om}
-
Y \Box\phi\)
\Bigr]
\nonumber\\
&=&
\frac{M_p^2\ga}{2}\DY\epsilon^{\al\bt\ga\lm}\epsilon^{\mu\nu\rho}{}_\lm \phi_{;\al}\phi_{;\mu}\phi_{;\bt;\nu}\phi_{;\ga;\rho}
\nonumber\\
&&
+ \frac{M_p^2}{2} 
\underbrace{
\ga\Dp \(
 \phi^{;\al} \phi{}_{;\al}{}_{;\om} \phi{}^{;\om}
\right.
}_{e_2}
\left.
-
Y \Box\phi\)\; .
\label{e1}
\ea
Let $\ga \Dp \equiv D_1 + Y D_{1,Y}$
and use $\phi^{;\al} Y_{;\al} = 2 \phi^{;\al} \phi{}_{;\al}{}_{;\om} \phi{}^{;\om}$,
one can integrate by parts the term $e_2$ in the above equation as
\be
e_2 = \frac 12 \(- D_1 Y \Box\phi - D_{1,\phi} Y^2\)\,.
\label{e2}
\ee
Inserting eqs.~(\ref{e1}) and (\ref{e2}) into eq.~(\ref{act-h}),
we can write the action for disformal gravity as
\ba
S_g &=& \int d^4 x \sqrt{-g}\biggl\{\,\,
G_2(\phi,Y)  + G_3(\phi,Y)\Box\phi+ G_4(\phi,Y) R 
\nonumber\\
&&\phantom{\int d^4 x \sqrt{-g}}~
 - 2 G_{4,Y}(\phi,Y) \Bigl[
\(\Box \phi\)^2
- \phi{}_{;\al}{}_{;\bt} \phi{}^{;\al}{}^{;\bt}\Bigr]
\nonumber\\
&&
\phantom{\,}
+ F_4(\phi,Y) \epsilon^{\al\bt\ga\lm}\epsilon^{\mu\nu\rho}{}_\lm \phi_{;\al}\phi_{;\mu}\phi_{;\bt;\nu}\phi_{;\ga;\rho}\,
\biggr\}\,,
\label{act-glpv-start}
\ea
where
\ba
G_2 &=&
- \frac{M_p^2}2 \frac Y2 \int (\ga \Dp)_{,\phi} dY\,,
\nonumber\\
G_3 &=&
- \frac{M_p^2}2 \[\frac 12 \int \ga \Dp dY
- \ga\Dp Y\]\,,
\nonumber\\
G_4 &=& \frac{M_p^2}{2\ga}\,,
\qquad
F_4 = \frac{M_p^2}2 \ga\DY\,.
\label{g24f4}
\ea
Setting $- D_1/2 = - \int \ga \Dp dY /2 \equiv 2 C_3 / M_p^2$,
we can write the action (\ref{act-glpv-start}) in the GLPV form as shown in eq.~(\ref{act-glpv})


\end{document}